\documentclass[sigconf, nonacm]{acmart}

\usepackage{tcolorbox}
\usepackage{amsmath}
\usepackage{algorithm}
\usepackage{algpseudocode}
\usepackage{amsfonts}
\usepackage{multirow}
\usepackage[normalem]{ulem}
\useunder{\uline}{\ul}{}
\usepackage{amsfonts}
\usepackage{color}

\newcommand\vldbdoi{XX.XX/XXX.XX}
\newcommand\vldbpages{XXX-XXX}
\newcommand\vldbvolume{18}
\newcommand\vldbissue{12}
\newcommand\vldbyear{2025}
\newcommand\vldbauthors{\authors}
\newcommand\vldbtitle{\shorttitle} 
\newcommand\vldbavailabilityurl{}
\newcommand\vldbpagestyle{empty} 

\begin{document}
\title{LEADRE: Multi-Faceted Knowledge Enhanced LLM Empowered Display Advertisement Recommender System }

\author{Fengxin Li}
\authornote{Work was done while Fengxin Li was intern at Tencent.}
\affiliation{%
  \institution{Renmin University of China}
  \country{}
}
\email{lifengxin@ruc.edu.cn}

\author{Yi Li}
\email{sincereli@tencent.com}
\author{Yue Liu}
\email{herculesliu@tencent.com}
\author{Chao Zhou}
\email{derekczhou@tencent.com}
\author{Yuan Wang}
\email{leoyuanwang@tencent.com}
\affiliation{%
  \institution{Tencent Inc.}
  \country{}
}

\author{Xiaoxiang Deng}
\email{reesedeng@tencent.com}
\author{Wei Xue}
\email{weixue@tencent.com}
\author{Dapeng Liu}
\email{rocliu@tencent.com}
\affiliation{%
  \institution{Tencent Inc.}
  \country{}
}

\author{Lei Xiao}
\email{shawnxiao@tencent.com}
\author{Haijie Gu}
\email{jerrickgu@tencent.com}
\author{Jie Jiang}
\email{zeus@tencent.com}
\affiliation{%
  \institution{Tencent Inc.}
  \country{}
}

\author{Hongyan Liu}
\authornote{Corresponding authors.}
\email{hyliu@tsinghua.edu.cn}
\affiliation{%
  \institution{Tsinghua University}
  \country{}
}

\author{Biao Qin}
\authornotemark[2]
\email{qinbiao@ruc.edu.cn}
\author{Jun He}
\authornotemark[2]
\email{hejun@ruc.edu.cn}
\affiliation{%
  \institution{Renmin University of China}
  \country{}
}

\begin{abstract}
Display advertising plays a crucial role in benefiting advertisers, publishers, and users. Traditional display advertising systems employ a multi-stage architecture comprising retrieval, coarse ranking, ranking, and re-ranking. However, conventional retrieval methods primarily rely on ID-based learning-to-rank mechanisms, often underutilizing the content information of ads, like ads' title, and description. This limitation reduces the ability to generate diverse and relevant recommendation lists.

To address this challenge, we propose leveraging the extensive world knowledge of large language models (LLMs). However, effectively integrating LLMs into advertising systems presents three key challenges: \textit{(i) How to accurately capture user interests}, \textit{(ii) How to bridge the knowledge gap between LLMs and advertising systems}, and \textit{(iii) How to efficiently deploy LLMs at scale}. To overcome these challenges, we introduce \textbf{LEADRE}—the \textbf{L}LM \textbf{E}mpowered Display \textbf{AD}vertisement \textbf{RE}commender system.
LEADRE consists of three core components. The \textbf{Intent-Aware Prompt Engineering} module introduces multi-faceted knowledge and constructs intent-aware \textit{<Prompt, Response>} pairs, fine-tuning LLMs to generate ads tailored to users' personal interests. The \textbf{Advertising-Specific Knowledge Alignment} module incorporates auxiliary fine-tuning tasks and Direct Preference Optimization (DPO) to align LLMs with advertising semantics and business objectives. The \textbf{Latency-Aware Model Deployment} module integrates a hybrid service framework that balances latency-tolerant and latency-sensitive service, ensuring seamless online deployment.

Extensive offline experiments validate the effectiveness of LEADRE, demonstrating significant improvements across multiple evaluation metrics. Furthermore, online A/B tests reveal a \textbf{1.57\%} and \textbf{1.17\%} increase in Gross Merchandise Value (GMV) for serviced users on WeChat Channels and Moments, respectively. LEADRE has been successfully deployed on both platforms, handling tens of billions of requests daily.

\end{abstract}

\maketitle

\pagestyle{\vldbpagestyle}
\begingroup\small\noindent\raggedright\textbf{PVLDB Reference Format:}\\
\vldbauthors. \vldbtitle. PVLDB, \vldbvolume(\vldbissue): \vldbpages, \vldbyear.\\
\href{https://doi.org/\vldbdoi}{doi:\vldbdoi}
\endgroup
\begingroup
\renewcommand\thefootnote{}\footnote{\noindent
This work is licensed under the Creative Commons BY-NC-ND 4.0 International License. Visit \url{https://creativecommons.org/licenses/by-nc-nd/4.0/} to view a copy of this license. For any use beyond those covered by this license, obtain permission by emailing \href{mailto:info@vldb.org}{info@vldb.org}. Copyright is held by the owner/author(s). Publication rights licensed to the VLDB Endowment. \\
\raggedright Proceedings of the VLDB Endowment, Vol. \vldbvolume, No. \vldbissue\ %
ISSN 2150-8097. \\
\href{https://doi.org/\vldbdoi}{doi:\vldbdoi} \\
}\addtocounter{footnote}{-1}\endgroup

\ifdefempty{\vldbavailabilityurl}{}{
\vspace{.3cm}
\begingroup\small\noindent\raggedright\textbf{PVLDB Artifact Availability:}\\
The source code, data, and/or other artifacts have been made available at \url{\vldbavailabilityurl}.
\endgroup
}

\section{Introduction}

Online display advertising plays a crucial role in facilitating targeted content delivery and meeting users' personal interest, benefiting advertisers, publishers, and users \cite{song2024multi, singh2023maximize}. Traditional display advertising systems employ a multi-stage architecture, including retrieval, coarse ranking, ranking, and re-ranking. The retrieval stage, which initiates the process, is critical for identifying user interests and mitigating the "information cocoon" effect by providing a diverse set of ad options \cite{li2021truncation, chen2022unified}.

Conventional retrieval methods primarily rely on ID-based learning-to-rank mechanisms to learn collaborative semantics for efficient ad filtering. However, these methods often underutilize ad content information, such as the title and description of the ads, which limits in generating diverse recommendations, particularly in scenarios with sparse user behaviors and long-tail ads \cite{wei2021contrastive, dong2020mamo}.

Recently, large language models (LLMs) have demonstrated remarkable capabilities in understanding, generalization, and reasoning by leveraging vast amounts of general knowledge \cite{nam2024using, zhang2024ucl, li2024planning}. Researchers have explored incorporating LLMs into recommender systems \cite{rajput2024recommender, wei2024llmrec} to enhance retrieval performance. However, applying LLMs in industrial-scale display advertising presents several critical challenges:

\textbf{(1) Capturing user interests in scenarios with implicit intents and sparse behaviors:}  
While LLMs exhibit strong capabilities in understanding user intent, display advertising often lacks explicit queries, making it difficult to infer user intent. Additionally, user interactions in the ad domain are typically sparse, necessitating effective utilization of limited data and supplementary behaviors to enhance commercial intent modeling.
\textbf{(2) Bridging the knowledge gap between LLMs and advertising systems:}  
Although LLMs excel in generating natural language responses based on general knowledge, display advertising requires the generation of relevant ads from a predefined inventory. It is essential to bridge the gap between the general knowledge of LLMs and the specific requirements of advertising systems. Furthermore, the generated ads must align with business objectives, requiring tailored fine-tuning strategies to optimize LLMs for advertising-specific goals.
\textbf{(3) Efficiently deploying LLMs in large-scale advertising systems serving billions of requests per day:}  
Integrating LLMs into an online advertising system imposes substantial computational demands, potentially conflicting with cost efficiency requirements. Thus, optimizing LLM deployment is essential to balance computational costs while maintaining scalability and real-time responsiveness.

To address these challenges, we propose \textbf{LEADRE}—a multi-faceted knowledge enhanced \textbf{L}LM \textbf{E}mpowered display \textbf{AD}vertisement \textbf{RE}commender system. LEADRE consists of three core components:

\textbf{(1) Intent-Aware Prompt Engineering:} 
This module designs intent-aware \textit{<Prompt, Response>} pairs that fine-tune LLMs to generate ads tailored to users' interests. To mitigate data sparsity, user behaviors from related content domains, such as micro-videos and news, are integrated. Furthermore, commercial intent is modeled by incorporating both \textbf{long-term interests} (from user profiles and historical behaviors) and \textbf{short-term interests} (from recent interactions) into the prompt.
\textbf{(2) Advertising-Specific Knowledge Alignment:} To bridge the semantic gap between natural language and advertising data, this module introduces auxiliary fine-tuning tasks. Additionally, Direct Preference Optimization (DPO) is applied to balance user intent with business objectives, ensuring that the generated ads have high commercial value.
\textbf{(3) Latency-Aware Model Deployment:} LEADRE is deployed using a hybrid architecture that integrates both latency-tolerant and latency-sensitive services. To further enhance computational efficiency, we optimize deployment using \textbf{TensorRT LLM Acceleration}.

We implement LEADRE using Hunyuan with 1B parameters and evaluate its performance through extensive offline and online experiments in Tencent's display advertising system. The offline results validate the contributions of the individual modules, demonstrating their effectiveness. In online A/B tests, LEADRE achieved a \textbf{1.57\%} increase in Gross Merchandise Value (GMV) on Tencent WeChat Channels and a \textbf{1.17\%} increase on Tencent WeChat Moments. Currently, LEADRE is deployed on both platforms, serving billions of users and processing tens of billions of requests per day.

The retrieved ads are further incorporated into the ranking phase as additional features. Specifically, on the user side, retrieved ads extend user interest representations, while on the item side, match scores between retrieved ads and target ads serve as new item-level features. These enhancements contribute to an additional \textbf{1.43\%} improvement in GMV on Tencent WeChat Channels.

The contributions of this work can be summarized as follows:

\begin{itemize}
    \item To the best of our knowledge, this is the first study to deploy LLMs in an online display advertising system. We introduce LEADRE, a novel LLM-based generative retrieval framework, and deploy it through a hybrid architecture that integrates both latency-tolerant and latency-sensitive service.
    \item LEADRE integrates Intent-Aware Prompt Engineering and Advertising-Specific Knowledge Alignment to ensure the generated ads are accurate, diverse, commercially valuable, and aligned with user interests.
    \item Extensive offline and online experiments demonstrate the effectiveness of our approach. We observe substantial improvements in offline metrics such as HitRatio, as well as in online metrics like GMV.
\end{itemize}

\section{Related Work}

\subsection{Sequential Recommendation}

Sequential recommendation leverages the user behavior sequence on items (ads) to predict the next item that user is likely to click or convert on \cite{xu2019survey, zhou2022filter}. Early approaches in sequential recommendation primarily focused on ID-based methods. These methods typically assign a unique ID to each item and employ sequential deep learning models, such as RNNs \cite{xu2019recurrent}, CNNs \cite{yan2019cosrec}, and Transformers \cite{kang2018self, sun2019bert4rec}, to learn sequence representations for next-item prediction. However, these approaches often struggle with issues related to cold start and data sparsity.
To address these challenges, subsequent research has incorporated additional feature information, including categorical features, numerical features, and graph structure features \cite{kumar2019predicting, yuan2021icai, chang2021sequential, liu2016context}. Some studies have also explored the integration of multi-modal information, such as item text descriptions and images \cite{li2023text, liu2021noninvasive, xie2022decoupled}. By leveraging this supplementary information, sequential recommendation methods can effectively mitigate cold start and data sparsity problems.

Moreover, some researchers have begun to adopt large language models (LLMs) to comprehend item features and infer user preferences, enabling direct recommendations without relying on traditional sequential recommendation methods \cite{bao2023tallrec, harte2023leveraging,wei2024llmrec, bao2024large}. In this paper, we introduce LLMs into a large-scale display advertising system to tackle challenges encountered in industrial scenarios, such as implicit user intent and high computational costs.

\subsection{Large Language Models-based Recommender}

Large Language Models (LLMs) have demonstrated remarkable capabilities in both understanding and generation tasks \cite{yao2024survey, huang2022large}, making them widely applicable across various domains, including document summarization \cite{koh2022empirical,zhang2024benchmarking}, conversational agents \cite{he2023large, wang2023enabling}, code completion \cite{guo2023longcoder, zhang2024llm}, and others \cite{zhu2023large, zhai2024large}. Recently, researchers have begun exploring the application of LLMs in recommender systems \cite{wu2024survey}. These efforts can be broadly categorized into two roles for LLMs: \textit{feature encoder} and \textit{ranker}.

As feature encoders, LLMs leverage their deep understanding of textual content to generate rich representations of users and items. Typically, user or item features are transformed into structured textual descriptions using pre-designed templates, which are then processed by LLMs to obtain feature embeddings \cite{ren2024representation, wei2024llmrec, bao2024large}. These embeddings can subsequently be utilized for downstream tasks such as candidate retrieval or ranking, improving personalization and recommendation quality.

As rankers, LLMs leverage their generative capabilities to predict user preferences based on historical behavior sequences. In this paradigm, user behavior sequences are represented as textual sequences, and LLMs are trained to predict the next likely item (or user action) based on context \cite{zhang2023chatgpt, harte2023leveraging}. To adapt LLMs effectively for ranking tasks, fine-tuning on domain-specific recommendation datasets is typically required, ensuring the models acquire relevant domain knowledge and effectively capture user preferences \cite{bao2023tallrec, harte2023leveraging, lin2024bridging}.

Despite these advancements, the direct application of LLMs in industrial display advertising remains challenging due to several factors, including implicit user intent, high computational costs, and the need for real-time inference. In this work, we present the first application of LLM-based generative retrieval in an industrial display advertising system. To address these challenges, we introduce intent-aware prompt engineering and ad-specific knowledge alignment, enabling LLMs to generate diverse, business-aligned, and user-tailored ad recommendations. Furthermore, we mitigate computational overhead by deploying LLMs in a hybrid architecture that integrates latency-tolerant and latency-sensitive services. This approach ensures that the benefits of LLM-based retrieval can be realized in a scalable and cost-efficient manner.

\subsection{Retrieval in Advertising Systems}

Display advertising systems typically follow a multi-stage architecture composed of multiple stages, including retrieval, corse-ranking, ranking, re-ranking, and others\cite{cui2020personalized,gharibshah2020deep}. The retrieval stage, positioned at the top of the funnel, plays a critical role in identifying user interests and generating a pool of candidate ads for further processing \cite{xie2022contrastive, ji2024neural}. Most advertising systems adopt ID-based learning to ranking mechanisms to recall all ads that users may find appealing. These methods often rely on a two-tower model, where one tower encodes user features and the other encodes item (ad) features \cite{cen2020controllable, huang2020embedding}.

Display advertising systems typically follow a multi-stage architecture consisting of retrieval, coarse ranking, ranking, re-ranking, and others \cite{cui2020personalized, gharibshah2020deep}. The retrieval stage, positioned at the entry of the process, is crucial for identifying user interests and generating an initial pool of candidate ads for subsequent ranking stages \cite{xie2022contrastive, ji2024neural}. 
Most advertising systems employ ID-based learning-to-rank mechanisms to retrieve relevant ads for users. These methods often utilize a two-tower model, where one tower encodes user features and the other encodes item (ad) features \cite{cen2020controllable, huang2020embedding}. While efficient, ID-based retrieval approaches exhibit several limitations: (1) Information Cocoon Effect: The reliance on past interactions can reinforce existing preferences, leading to a lack of novelty in recommendations. (2) Limited Content Awareness: ID-based approaches often fail to leverage rich ad content, missing potential signals for better personalization. (3) Reduced Diversity: The retrieved candidate set may be dominated by frequently interacted items, limiting exposure to new and diverse ads.

To address these challenges, we propose incorporating LLMs into the retrieval process. Unlike traditional ID-based methods, LLMs can analyze contextual information and generate a more diverse set of candidate ads by expanding retrieval beyond strict collaborative filtering signals. By leveraging their generative capabilities and semantic understanding, LLMs enhance retrieval diversity, improve content-awareness, and ultimately increase user engagement and advertising effectiveness.

\section{Preliminary}

\subsection{Problem Definition}

Let $\mathcal{U}$ and $\mathcal{A}$ denote the sets of users and advertisements (ads), respectively, within the target advertising system. Given the limited user interactions with ads, we introduce a content domain to enrich user behavior data. Let $\mathcal{C}$ denote the set of content items in the system. The sequence of user behaviors is represented as $\mathcal{S}_u = [i^1_u, i^2_u, \dots, i^L_u]$ in chronological order, where $u \in \mathcal{U}$, $i_u \in \mathcal{A} \cup \mathcal{C}$, and $L = |\mathcal{S}_u|$ is the length of the user's behavior sequence. Given this user behavior sequence, the goal of generative retrieval is to generate the next relevant ad, $a^{L+1}_u \in \mathcal{A}$, optimizing both user engagement and business objectives.

\begin{figure*}[t]
\centerline{\includegraphics[scale=0.6]{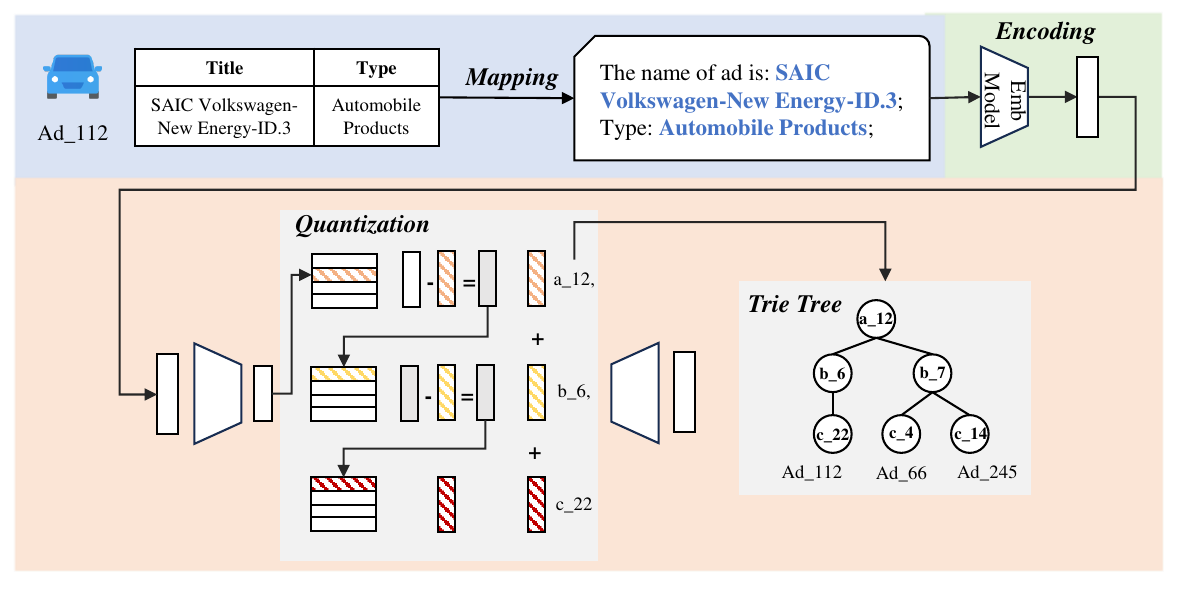}}
\caption{Overall Framework of Ads indexing in LEADRE.}
\label{fig_ad_id}
\end{figure*}

\subsection{Ads Indexing}

Traditional advertising systems typically index ads incrementally and learn dense embeddings for each ad \cite{wu2020sse, chang2021sequential}. However, this indexing lacks semantic information, creating a gap between the advertising system and human understanding. To bridge this gap, we introduce \textbf{Semantic IDs (S-IDs)} based on ad features, inspired by previous works on item indexing \cite{rajput2024recommender, zheng2024adapting, geng2022recommendation}.

To learn S-IDs, we first define a \textbf{feature mapping rule} in the form of a template string to generate text description of each ad, as follows:
\begin{tcolorbox}[colback=blue!2!white,leftrule=2.5mm,size=title]
    \emph{
    The name of the ad is \textbf{<ads\_name>}; The product type is \textbf{<ads\_type>}; The first-level category is \textbf{<first\_cat>}; The second-level category is \textbf{<second\_cate>}; The attributes include: \textbf{<basic\_att>}.
    }
\end{tcolorbox}

For each ad $a \in \mathcal{A}$, all its features are filled in above template string to generate a textual description $t_a$. For example:
\begin{tcolorbox}[colback=blue!2!white,leftrule=2.5mm,size=title]
    \emph{
    The name of the ad is \textbf{SAIC Volkswagen-New Energy-ID.3}; The product type is \textbf{Automobile Products}; The first-level category is \textbf{Automobile}; The second-level category is \textbf{SAIC Volkswagen·New Energy}; The attributes include: \textbf{automobile brand\_Volkswagen, automobile series\_Volkswagen ID.3}.
    }
\end{tcolorbox}

After obtaining the textual description $t_a$, we employ a pre-trained language model (e.g. Hunyuan \cite{sun2024hunyuanlargeopensourcemoemodel}, E5 \cite{wang2022text}) to \textbf{encode} $t_a$, bridging the gap between the advertising system and human understandable text. This step can be formulated as 
\begin{equation}
\mathbf{x}_a = f_{PLM}(t_a; \theta_{PLM})
\end{equation}
where $f_{PLM}$ represents the pre-trained language model parameterized by $\theta_{PLM}$, and $\mathbf{x}_a \in \mathbb{R}^{d_{PLM}}$ is the textual embedding of ad $a$, and $d_{PLM}$ denotes the output dimension of the pre-trained language model.

After that, we employ a Residual-Quantized Variational AutoEncoder (RQ-VAE) \cite{lee2022autoregressive} to generate concise semantic IDs for the ads. The RQ-VAE takes the ad’s  embedding $\mathbf{x}_a$ as input and uses an \textbf{encoder-quantization-decoder} mechanism to generate a list of discrete semantic tokens. 
In the encoding phase, the textual embedding $\mathbf{x}_a$ is encoded by an encoder, $f_{En}(\cdot;\theta_{En})$, to obtain a hidden embedding, formulated as $\hat{\mathbf{z}}_a = f_{En}(\mathbf{x}_a;\theta_{En})$, 
where $\hat{\mathbf{z}}_a \in \mathbb{R}^{d_{RQ}}$ denotes the hidden embedding, and $d_{RQ}$ is the hidden embedding's dimension.
During the quantization phase, we apply $M$ layers of residual vector quantization. For each layer $l$, we maintain a codebook $\mathcal{C}^l = \{\mathbf{e}_k^l\}_{k=1}^K$, where $\mathbf{e}_k^l \in \mathbb{R}^{d_{RQ}}$ and $K$ denotes the number of codes in the codebook. The quantization process is defined as follows:
\begin{align}
    c^l_a = &\text{argmin}_k || \mathbf{r}^l_a - \mathbf{e}_k^l ||_2^2  \\ 
    \mathbf{r}^{l+1}_a &= \mathbf{r}^{l}_a - \mathbf{e}_{c^l_a}^l  \\ 
    \mathbf{z}_a &= \sum_{l=1}^M \mathbf{e}_{c^l_a}^l 
\end{align}
where $\mathbf{r}^l_a \in \mathbb{R}^{d_{RQ}}$ is the $l$-th layer residual embedding of ad $a$, $\mathbf{r}^1_a = \hat{\mathbf{z}}_a$, $c^l_a$ is the index from the $l$-th codebook for ad $a$, and $\mathbf{z}_a \in \mathbb{R}^{d_{RQ}}$ is the final quantized embedding of ad $a$.
In the decoding phase, the quantized embedding $\mathbf{z}_a$ is used to reconstruct the original textual embedding $\mathbf{x}_a$ through a decoder $f_{De}(\cdot; \theta_{De})$,formulated as $\hat{\mathbf{x}}_a = f_{De}(\mathbf{z}_a; \theta_{De})$

To train the encoder $\theta_{En}$, decoder $\theta_{De}$, and codebooks $\{\mathcal{C}^l\}_{l=1}^M$, we introduce two loss components: the reconstruction loss and the quantization loss. The overall loss function for RQ-VAE is given by:
\begin{align}
    & \mathcal{L}_{recons} = \sum_a || \mathbf{x}_a - \hat{\mathbf{x}}_a ||_2^2  \\ 
    & \mathcal{L}_{quant} = \sum_a \sum_{l=1}^M \left( || \text{sg}[\mathbf{r}^{l}_a] - \mathbf{e}_{c^l_a}^l ||_2^2 + \beta_{quant} || \mathbf{r}^{l}_a - \text{sg}[\mathbf{e}_{c^l_a}^l] ||_2^2 \right)
\end{align}
where $\text{sg}[\cdot]$ is the stop-gradient operator, and $\beta_{quant}$ is a weight factor for the loss term.

\begin{algorithm}[t]
\caption{Trie-Tree Construction}
\label{algo_trie_tree}
\small{
\begin{algorithmic}[1]
\Require Set of all ads' S-IDs $\mathcal{C}_A = \{\mathbf{c}_1, \mathbf{c}_2, \dots, \mathbf{c}_a, \dots, \mathbf{c}_{|\mathcal{A}|} \}$, where each S-IDs sequence $\mathbf{c}_a = [c_a^1, c_a^2, \dots, c_a^l, \dots, c_a^{(M+1)}]$.
\Ensure Trie-Tree $T$ representing all ads in $\mathcal{A}$.

\State \textbf{Initialize:} Create a head node $h$ for the trie-tree.
\For{each S-IDs sequence $\mathbf{c}_a \in \mathcal{C}_A$}
    \State Set current node $cur \gets h$
    \For{each S-ID $c_a^l$ in $\mathbf{c}_a$}
        \If{$c_a^l$ \textbf{is not a child of} $c$}
            \State Create a new node $n$
            \State Add $n$ as a child of $cur$ with label $c_a^l$
        \EndIf
        \State Set $c \gets$ child of $cur$ labeled by $c_a^l$
    \EndFor
    \State Mark node $cur$ as the end of ad $a$
\EndFor
\State \Return Trie $T$
\end{algorithmic}
}
\end{algorithm}

This process ensures that ads are represented by discrete S-IDs, which allows the LLMs to generate valid ads from the predefined ads set during constrained decoding.
In the ad indexing process, each ad $a$ is represented by a list of semantic tokens, denoted as $\hat{\mathbf{c}}_a = [c_a^1, c_a^2, \dots, c_a^M]$. To address potential collisions where different ads map to the same S-IDs, we introduce an additional code to distinguish them \cite{rajput2024recommender}. Consequently, the S-IDs for each ad are updated to $ \mathbf{c}_a = [c_a^1, c_a^2, \dots, c_a^L, c_a^{(M+1)}] $, where $ c_a^{(M+1)} $ serves as the disambiguation code for ads with identical S-ID lists. To distinguish between different levels of semantic tokens, we use prefixes such as $ <a\_, b\_, c\_, \cdots>$ for various S-ID levels.

\subsection{Trie-Tree Construction}

We organize all ads using a trie-tree, where each path from the root to a leaf represents the S-IDs of an ad. This structure enables efficient ad organization and supports fast retrieval and prefix-based searches.
The \textbf{Trie-Tree Construction Algorithm} (Algorithm~\ref{algo_trie_tree}) builds the trie by sequentially adding each S-ID from an ad as a child node, creating new nodes as needed, and marking the end of each ad's S-ID sequence.

\textbf{Example:} Suppose we have the following ads, each represented by a sequence of S-IDs:
Ad\_66: $[a\_12, b\_7, c\_4]$; Ad\_245: $[a\_12, b\_7, c\_14]$; Ad\_112: $[a\_12, b\_6, c\_22]$.
We construct the trie-tree as follows:

\begin{enumerate}
    \item \textbf{Initialize} the trie with an empty root node.
    \item \textbf{Insert Ad\_66}: Add $a\_12$ as a child of root, then $b\_7$ under $a\_12$, and $c\_4$ under $b\_7$ (mark as ad end).
    \item \textbf{Insert Ad\_245}: $a\_12$ and $b\_7$ already exist; add $c\_14$ under $b\_7$ (mark as ad end).
    \item \textbf{Insert Ad\_112}: $a\_12$ exists; add $b\_6$ under $a\_12$, then $c\_22$ under $b\_6$ (mark as ad end).
\end{enumerate}
The resulting trie-tree (Figure~\ref{fig_ad_id}) allows for efficient ad retrieval and prefix-based searches, since ads sharing S-ID prefixes follow the same path from the root.







\begin{algorithm}[t]
\caption{Constrained Decoding Using Trie-Tree}
\label{algo_cons_decoding}
\small{
\begin{algorithmic}[1]
\Require Trie-tree $T$ representing all ads, Beam search size $B$, S-IDs list length $(M+1)$, Current token probability distribution $P_{LLM}(c^l | \mathbf{c}^{1:l-1};  \theta_{LLM})$ from the LLM.
\Ensure Generated ad set of ads $\mathcal{B}$.
\State Initialize generated ad set $\mathcal{B} \gets \{[h] \times B\}$, where $h$ is the head node of the trie-tree.
\For{$l = 1$ to $M+1$}
    \State Set current layer $cur\_layer$ as the $l$-th layer of trie-tree $T$.
    \State Initialize new generated set $\mathcal{B}' \gets \emptyset$.
    \For{each candidate list $\mathbf{c} \in \mathcal{B}$}
        \State Retrieve valid next S-IDs from the children of node $c^{l-1}$ in $cur\_layer$: $\mathcal{C}_{valid} \gets \text{children of}(c^{l-1})$.
        \For{each S-ID $c^l \in \mathcal{C}_{valid}$}
            \State Obtain the score for S-ID $c^i$ from the LLM: 
            \[
            P_{valid}(c^l) \gets P_{LLM}(c^l | \mathbf{c}^{1:l-1};  \theta_{LLM}).
            \]
            \State Append $c^l$ to the candidate list: $\mathbf{c} \gets \mathbf{c} \cup c^l$.
            \State Add the updated list $(\mathbf{c}, P_{valid}(c^l))$ to new set $\mathcal{B}'$.
        \EndFor
    \EndFor
    \State Update the generated ad set: $\mathcal{B} \gets \text{select\_top\_B}(\mathcal{B}', B)$.
\EndFor
\State \Return Generated ad set of ads $\mathcal{B}$.
\end{algorithmic}
}
\end{algorithm}

\subsection{Constrained Decoding} 
Unlike conventional LLM-based chatbots, generative retrieval in advertising requires generating a list of valid ads from a predefined set, ensuring both diversity and validity. To achieve this, we use constrained decoding with a trie-tree~\cite{hokamp2017lexically, hu2019improved, post2018fast}. This method ensures that generated ads are both likely according to the LLM and valid within the ad set. The \textbf{Constrained Decoding Algorithm} (Algorithm~\ref{algo_cons_decoding}) expands candidate S-ID lists layer by layer through the trie-tree, selecting the top $B$ candidates at each step based on their scores, until all layers or the maximum length is reached.

\textbf{Example:}  Suppose trie-tree is built from the following S-ID sequences:
Ad\_66: $[a\_12, b\_7, c\_4]$; Ad\_245: $[a\_12, b\_7, c\_14]$; Ad\_112: $[a\_12, b\_6, c\_22]$.
Using constrained decoding with a beam size $B=2$:
\begin{enumerate}
    \item \textbf{Initialization:}  
    Start with an empty beam set at the root.
    \item \textbf{First Layer:}  
    Valid first S-ID from the trie-tree is $a\_12$ with $P(a\_12)=0.6$.  
    Beam: $[a\_12]$ (score $0.6$).
    \item \textbf{Second Layer:}  
    For $[a\_12]$, valid next S-IDs are $b\_7$ and $b\_6$ with $P(b\_7)=0.5$, $P(b\_6)=0.4$:
    \begin{itemize}
        \item $[a\_12, b\_7]$ (score $0.6 \times 0.5 = 0.3$)
        \item $[a\_12, b\_6]$ (score $0.6 \times 0.4 = 0.24$)
    \end{itemize}
    Keep both as $B=2$.
    \item \textbf{Third Layer:}
    \begin{itemize}
        \item For $[a\_12, b\_6]$: valid next S-ID is $c\_22$ with $P(c\_22)=0.8$, yielding $[a\_12, b\_6, c\_22]$ (score $0.24 \times 0.8 = 0.192$).
        \item For $[a\_12, b\_7]$: valid next S-IDs are $c\_4$ and $c\_14$ with $P(c\_4)=0.8$, $P(c\_14)=0.4$:
        \begin{itemize}
            \item $[a\_12, b\_7, c\_4]$ (score $0.3 \times 0.8 = 0.24$)
            \item $[a\_12, b\_7, c\_14]$ (score $0.3 \times 0.4 = 0.12$)
        \end{itemize}
    \end{itemize}
    After scoring, keep the top $B=2$: $[a\_12, b\_7, c\_4]$ and $[a\_12, b\_6, c\_22]$.
\end{enumerate}
The algorithm returns the final beam set of S-ID sequences, ensuring that the generated ads are valid according to the trie-tree and have the highest predicted probabilities.


\begin{figure*}[t]
\centerline{\includegraphics[scale=0.55]{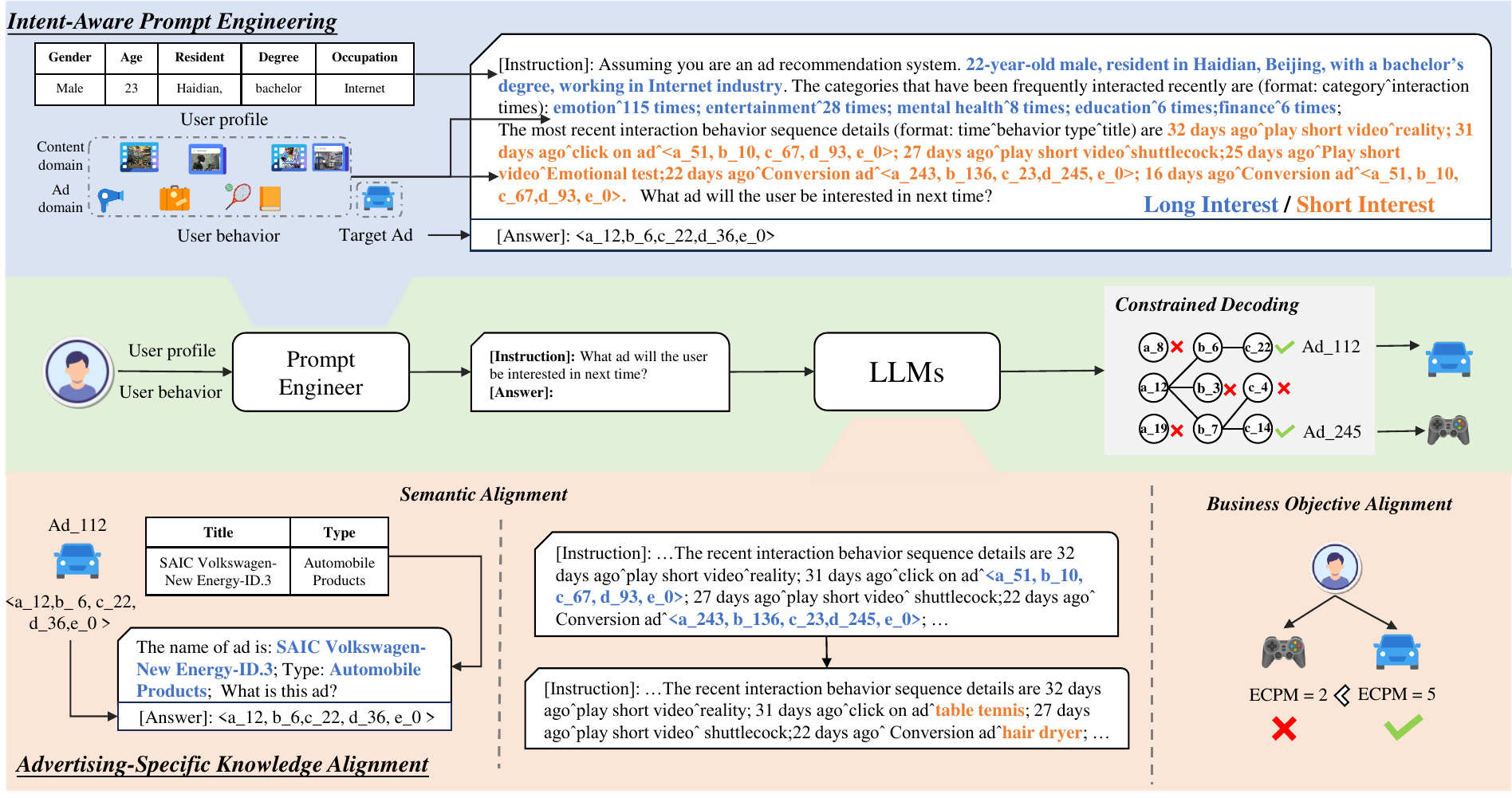}}
\caption{Overall Framework of Ads Constrained Generation Module and LLM Fine-tuning Module.}
\label{fig_framework}
\end{figure*}

\section{Method}

In this section, we present a detailed explanation of multi-faceted knowledge enhanced \textbf{L}LM \textbf{E}mpowered display \textbf{AD}vertisement \textbf{RE}commender system (LEADRE), which integrates large language models (LLMs) into our display advertising system.



\subsection{Overview}

LEADRE effectively aligns user intent with business objectives, improving the generation of relevant and high-value ads in advertising environments. It's framework consists of three novel modules:

\textbf{(1) Intent-Aware Prompt Engineering:} This module designs intent-aware \textit{<Prompt, Response>} pairs that incorporate user profiles and behavior sequences, serving as a fine-tuning corpus for LLMs to learn user interests effectively.

\textbf{(2) Advertising-Specific Knowledge Alignment:} This module integrates auxiliary fine-tuning tasks and Direct Preference Optimization (DPO) to bridge the knowledge gap between LLMs and the advertising system, enhancing ad generation quality.

\textbf{(3) Latency-Aware Model Deployment:} This module combines both latency-tolerant and latency-sensitive service systems to ensure scalable and real-time performance during LLM deployment.

\subsection{Intent-Aware Prompt Engineering}
\label{sec_llm_fine_tune}
To accurately capture user intent and predict the next relevant ad, the LLM should be fine-tuned with proper corpus first. We build textual \textit{<Prompt, Response>} pairs with user behavior sequences and ad descriptions as the corpus. Unlike search advertising \cite{rajput2024recommender}, where user queries explicitly express intent and could be used as the prompt, display advertising operates in a low-intent environment where explicit queries are absent and user behavior is often sparse. As a result, the prompt must effectively leverage  available user data within the advertising system and supplement it with user data outside of the advertising system, such as user behaviors in the content domain, to adequately capture commercial intent. Considering the affordable prompt length is quite limited, user data should be encoded concisely enough. Below, we detail the prompt engineering process.

\subsubsection{Prompt Components}
To enable LLMs to capture user commercial intent effectively, we incorporate both \textbf{long-term interests} (derived from user profiles and historical behaviors) and \textbf{short-term interests} (based on recent behaviors) into prompt design. 
Short-term interests model dynamic, immediate preferences. For example, a user planning to purchase a phone would likely click phone or phone accessory ads within a brief period. However, such interests are inherently noisy and transient (e.g., after purchasing the phone, the user would cease clicking similar ads).
Conversely, long-term interests capture stable, enduring preferences. For example, a student user would consistently engage with educational coaching ads over extended periods. Specifically, The prompt is assembled with the following key components:

    \textbf{(1) Task Instruction}: This part describes the task for the LLM to perform. Furthermore, it provides a learnable token for all fine-tuning tasks, similar to Soft Prompt formats \cite{qin2021learning, vu2021spot}, helping the LLM recognize instructions related to ad generation and retrieval. The template of the task description is shown below:
     
    \begin{tcolorbox}[colback=blue!2!white,leftrule=2.5mm,size=title]
    \emph{
    The following is an instruction describing a task. Please give a response to complete this request appropriately.}
    
    \emph{
    [Instruction]:\textbf{<Learnable Token>}Assuming you are an ad recommender system, \textbf{<Prompt>}, what ad will the user be interested in next time?
    }
    \end{tcolorbox}
    
    \textbf{(2) User Profiles}: This part provides demographical information about the user, such as age, gender, and region. The template is shown below:

    \begin{tcolorbox}[colback=blue!2!white,leftrule=2.5mm,size=title]
    \emph{
    \textbf{<age>}\textbf{<sex>}, resident in \textbf{<residence>}, with a \textbf{<education\_level>} degree, working in \textbf{<occupation>}, with a \textbf{<consumption\_level>}
    }
    \end{tcolorbox}

    \textbf{(3) User Interest Summary}: This part summarizes the user's long-term positive and negative feedbacks on commercial interest categories and standard product units (SPUs) across the ad and content domains. The template is shown below:

    \begin{tcolorbox}[colback=blue!2!white,leftrule=2.5mm,size=title]
    \emph{
    The categories that have been frequently interacted recently are (format: category\^{}interaction times): \textbf{<category\_1>}\^{}\textbf{<count\_1>} times; \textbf{<category\_2>}\^{}\textbf{<count\_2>} times; \textbf{<category\_3>}\^{}\textbf{<count\_3>} times; 
    }
    \end{tcolorbox}

    \textbf{(4)User Ad Domain Behavior Sequence}: This part expresses the user's short-term behaviors in the ad domain. To filter the noise signals in the sequence, only positive user actions like clicks and conversions are considered. As discussed in Section 3.1, each ad is represented by a list of semantic tokens denoted as Semantic-IDs (S-IDs). To distinguish different level semantic tokens, we use <a\_ , b\_, c\_, $\cdots$ > as prefix of different S-IDs level.

    \textbf{(5) User Content Domain Behavior Sequence}: This part captures the user's short-term behaviors in the content domain. Each interaction is described in terms of the commercial category and SPU. Similarly, only positive actions like watching full videos and searching are considered. The ad and content domain behaviors are merged and presented in chronological order. The template is shown below:

    \begin{tcolorbox}[colback=blue!2!white,leftrule=2.5mm,size=title]
    \emph{
    The most recent interaction behavior sequence details (format: time\^{}behavior type\^{}title) are \textbf{<time\_1>} days ago\^{}\textbf{<type\_1>}\^{} \textbf{<title\_1>}/\textbf{<SIDs\_1>}; \textbf{<time\_2>} days ago\^{}\textbf{<type\_2>}\^{}\textbf{<title\_2>}/\textbf{<SIDs\_2>}; \textbf{<time\_3>} days ago \^{} \textbf{<type\_3>} \^{} \textbf{<title\_3>}/\textbf{<SIDs\_3>}
    }
    \end{tcolorbox}

By concatenating the previous components, we obtain a complete prompt. An example prompt is shown below:
\begin{tcolorbox}[colback=blue!2!white,leftrule=2.5mm,size=title]
    \emph{
    The following is an instruction describing a task. Please give a response to complete this request appropriately.  }  
    \emph{
    \textbf{22-year-old male}, resident in \textbf{Haidian, Beijing}, with a \textbf{bachelor}'s degree, working in \textbf{Internet industry}, with a \textbf{medium} consumption level.
    The categories that have been frequently interacted recently are (format: category\^{}interaction times):  \textbf{emotion\^{}115 times; entertainment\^{}28 times};  \textbf{mental health\^{}8 times};  \textbf{education\^{}6 times}; \textbf{finance\^{}6 times}; }
    \emph{
    The most recent interaction behavior sequence details (format: time\^{}behavior type\^{}title) are \textbf{32 days ago\^{}play short video\^{}reality}; \textbf{31 days ago\^{}click on ad\^{}<a\_51, b\_10, c\_67, d\_93, e\_0>}; \textbf{27 days ago\^{}play short video\^{}shuttlecock}; \textbf{25 days ago\^{}Play short video\^{}Emotional/psychological age test};\textbf{22 days ago\^{}Conversion ad\^{}<a\_243, b\_136, c\_23, d\_245, e\_0>};\textbf{19 days ago\^{}Click ad\^{}<a\_164, b\_243, c\_38, d\_88, e\_0>};\textbf{16 days ago\^{}Conversion ad\^{}<a\_51, b\_10, c\_67, d\_93, e\_0>}.
    what ad will the user be interested in next time?
    }
\end{tcolorbox}

For the $Response$, the last click or conversion ad in the user behavior sequence is used as a supervisory signal for fine-tuning the LLM. A sample response is shown below:
\begin{tcolorbox}[colback=blue!2!white,leftrule=2.5mm,size=title]
    \emph{
    [Response]: \textbf{<a\_122, b\_28, c\_35, d\_15, e\_0>}
    }
\end{tcolorbox}

\subsubsection{Prompt Augmentation}

We develop a series of prompt augmentation strategies to enhance the fine-tuning process. These strategies are outlined below:

\textbf{Multiple Prompt Templates:}  
We introduce a variety of prompt templates to map a single user behavior sequence into multiple prompts. By altering the arrangement of sequence components and varying the instruction descriptions, these templates facilitate a broader understanding of user intentions. This diversification allows the model to capture different aspects of user intent, enriching the generation process.

\textbf{User Profile Reordering:}  
To prevent LLMs from simply memorizing specific token sequences or orders, we propose a user profile reordering method. It reshuffles user profile descriptions, encouraging the model to focus on the semantic meaning rather than the fixed order of inputs.

\textbf{Ad Positive Interaction Reuse:}  
Inspired by auto-regressive mechanisms, We augment the original user behavior sequence into multiple prompt samples by reusing positive interaction, enhancing the model’s ability to generalize from limited interaction data. For example, Assuming that user behavior sequence is denoted by $\mathcal{S}_{u} = [c^1_u, a^2_u, a^3_u,c^4_u ,a^5_u]$, where $c^1_u, c^4_u \in \mathcal{C}$ denote content domain items and $a^2_u, a^3_u, a^5_u \in \mathcal{A}$ denote ads. We augment the original sequence into prompt samples: <$[c^1_u, a^2_u, a^3_u, c^4_u]$, $a^5_u$>, <$[c^1_u, a^2_u ]$, $a^3_u$>, <$[c^1_u]$, $a^2_u$>.

\subsection{Advertising-Specific Knowledge Alignment}
The advertising system poses a great gap with LLMs, which hurts the ad generation ability of LLMs. To bridge this gap, we conduct semantic alignment by auxiliary tuning tasks to align the ad's Semantic-IDs (S-IDs) with the LLM, and business objectiveness alignment by Direct Preference Optimization (DPO) \cite{rafailov2024direct, yuan2024rrhf} to encourage the generation of ads with higher Effective Cost Per Mille (ECPM).

\subsubsection{Semantic Alignment}
Since LLM lacks prior knowledge of ads and their S-IDs, direct training on the main task encourages rote memorization rather than genuine understanding. This likes teaching calculus to students without foundational mathematics, which often results in solution memorization instead of conceptual comprehension. To address this, we introduce auxiliary tasks to guide LLMs learn the basic understand of advertising system and S-IDs of ads\cite{zheng2024adapting}. The auxiliary tasks consist of two components: an explicit alignment task and an implicit alignment task. 

\textbf{The explicit alignment task} enables LLM to acquire S-ID knowledge by predicting S-IDs from detailed ad descriptions. This task directly aligns the language model with the ad system by making the LLMs understand how textual descriptions map to the corresponding ad S-IDs. A sample \textit{prompt-response} pair for this task is shown below:

\begin{tcolorbox}[colback=blue!2!white,leftrule=2.5mm,size=title]
    \emph{
    [Prompt]: Given the ad's detailed description "\textbf{The name of the ad is SAIC Volkswagen-New Energy-ID.3; The product type is Automobile Products; The first-level category is Automobile; The second-level category is SAIC Volkswagen·New Energy; The attributes include: automobile brand\_Volkswagen, automobile series\_Volkswagen ID.3.}", what is the corresponding ad? 
    }
    
    \emph{
    [Response]: \textbf{<a\_12, b\_22, c\_50, d\_25, e\_0>}
    }
\end{tcolorbox}

In this example, LLM is provided with a detailed textual description of an ad, including the product's name, type, brand, and category. The model's task is to map this description to the corresponding S-IDs, which are internal identifiers used by the ad system. By adopting this task, LLM will have a better understanding about the S-IDs and ads' feature, which plays a basic ability of generating ads.

\textbf{The implicit alignment task} develops advertising knowledge within user interaction contexts.
This task builds on the Next Ad Generation mechanism described in the main task but replaces the S-IDs in the user's ad interaction history with the ad descriptions. Instead of predicting the next ad based on a sequence of S-IDs (as in the main task), the model now predicts the next ad using textual descriptions of the ads. A sample \textit{<Prompt, Response>} pair for this task is shown below:

\begin{tcolorbox}[colback=blue!2!white,leftrule=2.5mm,size=title]
    \emph{
    [Prompt]: The following is an instruction describing a task. Please give a response to complete this request appropriately.  $\cdots$
    The most recent interaction behavior sequence details (format: time\^{}behavior type\^{}title) are \textbf{32 days ago\^{}play short video\^{}reality}; \textbf{31 days ago\^{}click on ad\^{}Bali 7-Day Tour}; \textbf{27 days ago\^{}play short video\^{}shuttlecock}; \textbf{25 days ago\^{}Play short video\^{}Emotional/psychological age test};\textbf{22 days ago\^{}Conversion ad\^{}psychological test};\textbf{19 days ago\^{}Click ad\^{}Educational Counseling};\textbf{16 days ago\^{}Conversion ad\^{}Tarot reading}.
    What ad will the user be interested in next time?
    }

    \emph{
    [Response]: \textbf{<a\_122, b\_28, c\_35, d\_15, e\_0>}}
\end{tcolorbox}

This subtle modification enhances the LLMs' ability to associate ad descriptions with user behaviors, further aligning the model with the advertising system in a more implicit manner.

\textbf{Task Order:} The tasks are conducted in the following order: explicit alignment task $\rightarrow$ implicit alignment task $\rightarrow$ main task. It is essential to prioritize alignment tasks, both explicit and implicit, before addressing the main task. These alignment tasks, particularly the explicit alignment task, serve as crucial preparatory steps that equip the LLMs to understand the mapping between textual ad descriptions and the S-IDs used by the ad system. 
By completing these tasks first, the model gains a deeper comprehension of the structural and semantic aspects of the ad domain, ensuring that it is properly aligned with the system's internal representations. This alignment process is not merely a supplementary step but a foundational one, directly influencing the model's effectiveness in the primary task. In essence, the explicit and implicit alignment tasks establish the groundwork necessary for the model's success in predicting the next ad, making them indispensable for optimal performance in the main task.

\begin{figure*}[t]
\centerline{\includegraphics[scale=0.6]{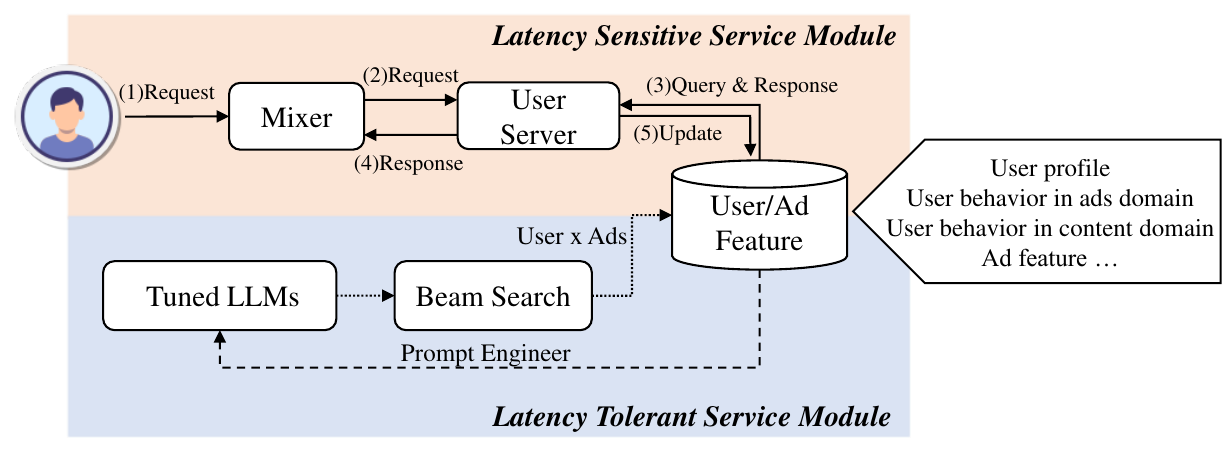}}
\caption{Framework of Latency-Aware Model Deployment.}
\label{fig_deploy}
\end{figure*}

\subsubsection{Business Objectiveness Alignment}
Since LLMs inherently lack business context, they may treat all advertisements as having equivalent value. However, in advertising systems, each ad possesses distinct business value that directly impacts system revenue. When tuned exclusively using prior tasks, the generated ads often demonstrate limited commercial value, contributing minimally to both cost and Gross Merchandise Volume (GMV). To address this limitation, we employ \textbf{Direct Preference Optimization (DPO)} to encourage generating ads with higher business value.

Since LLMs inherently lack business context, they may treat all advertisements as having equivalent value. However, in advertising systems, each ad possesses distinct business value that directly impacts system revenue. When tuned exclusively using prior tasks, the generated ads often demonstrate limited commercial value, contributing minimally to both cost and Gross Merchandise Volume (GMV). To address this limitation, we employ \textbf{Direct Preference Optimization (DPO)} to encourage generating ads with higher business value.

As defined in Section 3.1, for a specific user $u$, we aim to predict the next interacted ad $a^{L+1}_u$ based on user behavior sequence $\mathcal{S}_u$. When considering two equally appealing ads $a_1$ and $a_2$ where $a_1$ holds greater business value, recommending $a_1$ is preferable due to its higher revenue potential. We therefore implement DPO to increase the probability of generating high-value ads ($a_h$) while reducing probability for lower-value ads ($a_l$). DPO preference pairs $\langle u, a_h, a_l \rangle$ are constructed using the Expected Cost Per Mille (ECPM) metric from the ranking phase.

Specifically, for a given user $u$, we select two potential ads, $a_1$ and $a_2$, that are likely to result in a click or conversion. We then calculate the ECPM for each ad. The ad with the higher ECPM is considered the high-value ad $a_h$, while the ad with the lower ECPM is considered the low-value ad $a_l$. 
The training loss associated with DPO can be defined as:
\begin{equation}
    \mathcal{L}_{DPO} = 
    - \mathbb{E}_{(u, a_h, a_l)\sim \mathcal{D}}
    \left[\log \sigma\left(
    \beta \frac{\pi_{\theta}(a_h|u)}{\pi_{ref}(a_h|u)} -\beta \frac{\pi_{\theta}(a_l|u)}{\pi_{ref}(a_l|u)}
    \right)\right]  \notag
\end{equation}
where $\mathcal{D}$ is the set of collected DPO training triplets, $\pi_{\theta}$ is the LLM tuned by DPO loss, $\pi_{ref}$ is the LLM tuned by primary and auxiliary tasks, $\sigma(\cdot)$ is the sigmoid function, and $\beta$ is a hyper-parameter. This loss function aims to maximize the likelihood of generating high-value ads while minimizing the likelihood of low-value ads, thereby aligning the LLM's output more closely with business objectives.

\subsection{Latency-Aware Model Deployment}

\label{sec_deploy}

In this section, we detail the deployment framework and engineering techniques for LLM inference used to ensure efficient and responsive ad generation.

\subsubsection{Deployment Framework}
Typical LLMs exhibit high latency, which is inadequate for real-time display advertising services. To address this, we implement a hybrid system by integrating latency tolerant service module and latency sensitive service module \cite{li2021truncation, chen2022unified}. The deployment consists of the following components:

\textbf{(1) Inference: } 

\textit{Latency Sensitive Service Module :} This component is activated when a user makes a request. A Mixer receives the user request and queries the User Server for the ad retrieval list. The User Server retrieves the pre-computed retrieval list from the user feature database and sends it to the Mixer and further responds to the user. After that, the user behaviors update and ad feature update will be stored in the user feature database and ad feature database.

\textit{Latency Tolerant Service Module:} This module adopts nearline computing for LLM inference and is triggered after a user request. This component receives the fine-tuned LLM and the Trie-Tree from the offline stage for inference. It inquiries users' recent behavior to construct prompt samples and generates a list of ads for the users. The generated ad list is stored in the user feature database and made available for online service. 

Even for nearline inference, GPU resources are relatively scarce and cannot be allocated to every request. To address this, we introduce the \textbf{Adaptive Resource Distribution Strategy}. This strategy categorizes users into 25  user groups according to Average Revenue Per User (ARPU) value, prioritizing GPU inference for high-value users.

\textbf{(2) Training:} The training stage is responsible for LLM fine-tuning, and it is updated daily. The process begins with engineering samples from the user feature database to extract ad features to construct \textit{<Prompt, Response>} pairs for LLM fine-tuning. The \textit{<Prompt, Response>} pairs are utilized to fine-tune the LLM, enabling it to generate ads effectively. Once the training is complete, the fine-tuned LLM is deployed to the latency tolerant service module.

\subsubsection{TensorRT LLM Acceleration}
To further enhance the computational efficiency of nearline LLM inference, we employ TensorRT \footnote{https://github.com/NVIDIA/TensorRT-LLM/tree/release/0.5.0} for inference acceleration. This approach leverages several optimization techniques to significantly reduce latency and improve performance, particularly in advertising scenarios:

\noindent 1) \textbf{TensorRT-LLM Kernel Optimization}

 \textit{Softmax Kernel Optimization:}  
     The softmax kernel, which accounts for over 80\% of decoding time, is optimized by using 'float4' for vectorized memory access, reducing bandwidth usage and doubling kernel performance. This reduces end-to-end time by 5\%. Additionally, for the prefill step, we introduced a `ComputeMode` to calculate softmax only for beam 0, improving prefill performance by 3-4 times.
    
\textit{Finalize Kernel Optimization:}  
    In beam search, the finalize kernel recursively obtains token IDs. Initially, it used a single thread, underutilizing GPU resources. By assigning a block per beam width, performance improved by 50 times, reducing latency by 10\%.

\noindent 2) \textbf{TensorRT-LLM Quantization:} LLM inference generally involves two stages: prefill and generate. Prefill is typically the computational bottleneck, while the generate stage is more memory-bound. However, in advertising scenarios with a large beam width and batching functionality supported by the inference engine, the generate stage also becomes a computational bottleneck. Therefore, we adopt quantization schemes that focus on activation values, such as smooth quantization (`w8a8c8`) and FP8 quantization (`w8a8c8`), to mitigate these bottlenecks.
    
 \textit{Smooth Quantization (`w8a8c8`):}  
    We apply smooth quantization (int8 precision for weights and activations), doubling the computational power for operations like matrix multiplication. This results in a 50\% performance improvement in search advertising.
    
 \textit{FP8 Quantization (`w8a8c8`):}  
    FP8 quantization offers a simpler alternative to smooth quantization and achieves similar performance on H20 GPUs without sacrificing precision.

\noindent 3) \textbf{Proxy Load Balancing for Multi-GPU Utilization:}  
Previous optimizations, such as proxy global load balancing, mainly focused on improving the performance of individual GPUs. To enhance the utilization of multiple GPUs, we optimized the scheduling strategy to better distribute workloads across the GPU cluster. In our system, multiple proxies interact with multiple TensorRT-LLM instances. Without coordination, multiple proxies may send requests to the same TensorRT-LLM instance, leading to uneven load distribution.

TensorRT-LLM is highly sensitive to queries per second (QPS). If one instance processes even a couple more requests than others, it can become the bottleneck for the entire cluster. To mitigate this, we introduced Redis-based global counters to evenly distribute requests across all TensorRT-LLM instances, ensuring balanced load distribution. As a result, the TensorRT-LLM cluster now achieves over 90\% of its theoretical maximum load capacity.

\section{Experiment}

In this section, we verify the effectiveness of LEADRE by conducting extensive offline and online experiments.

\subsection{Experiment Setup}

\begin{table}[t]
\caption{ Comparison of performances under different components of prompt. "w.o. content" variant removes the content domain behaviors from the prompt, retaining only the ad domain behaviors, and "w.o. summary" removes the user interest summary from the prompt. }
\label{tab_ab_prompt}
\scalebox{0.85}{
\begin{tabular}{c|cccccc}
\toprule
 Variants                       & HR@1            & HR@4            & HR@8            & NDCG@4          & NDCG@8          \\ \hline
LEADRE                           & \textbf{0.0764} & \textbf{0.1567} & \textbf{0.2021} & \textbf{0.1199} & \textbf{0.1360} \\ \hline
\multirow{2}{*}{w.o. content}   & 0.0660          & 0.1343          & 0.1773          & 0.1029          & 0.1181          \\
                                & -13.60\%        & -14.28\%        & -12.26\%        & -14.19\%        & -13.15\%        \\ \hline
\multirow{2}{*}{w.o. summary}   & 0.0760          & 0.1543          & 0.2043          & 0.1181          & 0.1358          \\
                                & -0.62\%         & -1.54\%         & +1.08\%         & -1.52\%         & -0.18\%         \\ \bottomrule
\end{tabular}}
\end{table}

\begin{figure}[]
\centerline{\includegraphics[scale=0.45]{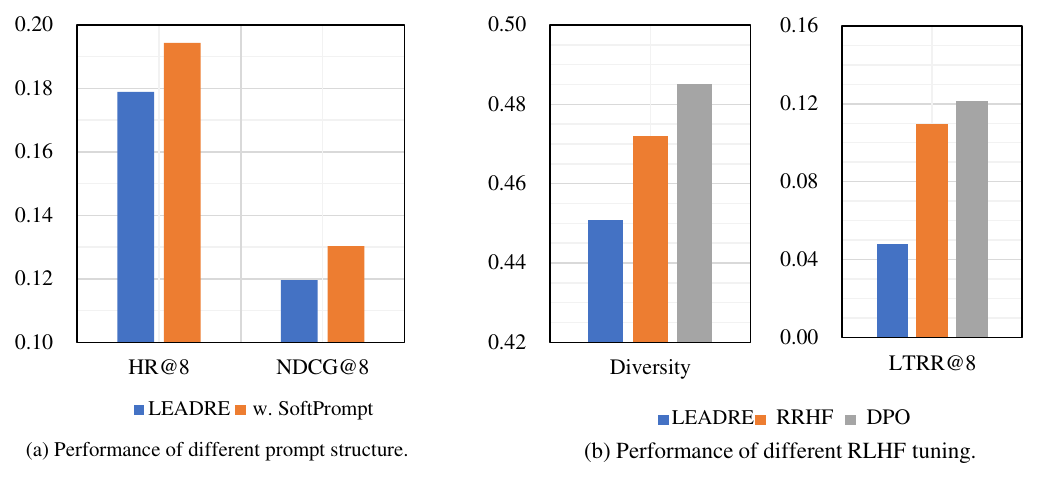}}
\caption{Ablation Studies Results on (a) Soft Prompt and (b) DPO tuning.}
\label{fig_ab_dpo}
\end{figure}

\subsubsection{Dataset}
We conducted all experiments using industrial-scale datasets on display advertising of Tencent WeChat Channels, as public datasets were deemed unsuitable due to gaps in applicability to our serving system and significant discrepancies with our internal models. The experiment process is suitable for all industrial display advertising systems. 

We created user behavior sequences in chronological order and applied the "leave-one-out" strategy for dataset splitting \cite{harte2023leveraging, wu2020sse, chang2021sequential}. Specifically, the last ad interaction in each user's sequence was designated as the test set, while the remaining interactions were used for training. The maximum user behavior window was set to 90 days and the maximum prompt token length was set to 2096, covering user behavior in both the ads and content domains.


\subsubsection{Evaluation Metric}
We used the Hit Ratio (HR@K) and Normalized Discounted Cumulative Gain (NDCG@K) as the primary metrics to evaluate the offline predictive performance of the models. Hit Ratio measures whether the model successfully retrieves the clicked or converted ads for users. NDCG accounts for both the relevance of items and their positions in the ranked list. 

\subsubsection{Implementation Details}
The LEADRE is implemented on the Hunyuan with 1B parameters \cite{sun2024hunyuanlargeopensourcemoemodel}. Training was conducted on 16 A100Pro GPUs, while inference was performed on hundreds of L40S GPUs.
For the Ads indexing, we employed the Hunyuan to encode ad features. The number of residual quantization steps was set to 4, with each layer containing 1024 code vectors, each having a dimension of 8. The length of the S-ID sequence was set to 5. 
For LLM fine-tuning, we employed the AdamW optimizer, setting the learning rate to 6e-5, weight decay to 0.9, and a minimum learning rate of 6e-6. By leveraging data parallelism and gradient accumulation, the batch size was set to 64. To avoid overfitting, training was performed for 3 epochs.


\subsection{Offline Performance}

\begin{table}[t]
\caption{Comparison of performances under different tuning tasks and order. "Main" denotes the main task (Next Ad Generation), "EX" denotes the explicit alignment task, "IM" denotes the implicit alignment task, and "mix" denotes mix the tuning $<Prompt, Response>$ pairs and random sample pairs to tune the LLM.  }
\label{tab_ab_task}

\scalebox{0.85}{
\begin{tabular}{c|ccccc}
\toprule
Tasks  and Order                            & HR@1            & HR@4            & HR@8            & NDCG@4          & NDCG@8          \\ \hline
EX $\rightarrow$   IM $\rightarrow$ Main & \textbf{0.0780} & \textbf{0.1610} & \textbf{0.2062} & \textbf{0.1232} & \textbf{0.1391} \\ \hline
\multirow{2}{*}{EX$\rightarrow$ Main}   & 0.0767          & 0.1566          & 0.1998          & 0.1201          & 0.1354          \\
                                        & -1.71\%         & -2.74\%         & -3.09\%         & -2.50\%         & -2.67\%         \\ \hline
\multirow{2}{*}{Main+IM+EX mix}         & 0.0745          & 0.1523          & 0.2009          & 0.1165          & 0.1337          \\
                                        & -4.60\%         & -5.43\%         & -2.55\%         & -5.39\%         & -3.90\%         \\ \hline
\multirow{2}{*}{Main+EX mix}            & 0.0766          & 0.1516          & 0.1948          & 0.1174          & 0.1327          \\
                                        & -1.91\%         & -5.84\%         & -5.49\%         & -4.64\%         & -4.59\%         \\ \hline
\multirow{2}{*}{Main only}              & 0.0707          & 0.1444          & 0.1879          & 0.1107          & 0.1260          \\
                                        & -9.42\%         & -10.32\%        & -8.87\%         & -10.16\%        & -9.42\%         \\ \bottomrule

\end{tabular}}
\end{table}

\subsubsection{Effectiveness of Prompt Components}
We perform ablation studies on the different components of the prompt in the main task (Section~\ref{sec_llm_fine_tune}) to assess their individual contributions.

 \textit{Content Domain Behaviors and Interest Summary:}  
To evaluate the impact of content domain behaviors and user interest summary, we design two variants: "w.o. content" and "w.o. summary". The "w.o. content" variant removes the content domain behaviors from the prompt, retaining only the ad domain behaviors, while "w.o. summary" removes the user interest summary from the prompt. As shown in Table~\ref{tab_ab_prompt}, both variants lead to a performance drop across all metrics. This demonstrates the effectiveness of including both content domain behaviors and the user interest summary in capturing user preferences and intent.

 \textit{Soft Prompt:}  
To investigate the effectiveness of adding an instruction-based task description, we introduce a learnable token at the beginning of the prompt, referred to as "w. SoftPrompt". The results, depicted in Figure~\ref{fig_ab_dpo}(a), show a performance improvement after incorporating the learnable token, confirming the positive impact of the SoftPrompt on the task.

\subsubsection{Effectiveness of Tuning Tasks and Their Order}
To study the effect of various tuning tasks and their order, we design several tuning strategies: "EX $\rightarrow$ IM $\rightarrow$ Main", "EX $\rightarrow$ Main", "Main+IM+EX mix", "Main+EX mix", and "Main only". The "EX $\rightarrow$ IM $\rightarrow$ Main" strategy refers to tuning the LLM sequentially on the explicit alignment task, implicit alignment task, and then the main task. The "EX $\rightarrow$ Main" strategy skips the implicit task and directly tunes on the explicit task followed by the main task. "Main+IM+EX mix" and "Main+EX mix" represent mixed-tuning strategies where pairs are randomly sampled and tuned together. "Main only" refers to tuning solely on the main task, without alignment tasks.

Table~\ref{tab_ab_task} presents the tuning results, leading to the following observations:
(1) The explicit and implicit alignment tasks enhance the performance of the main task by bridging the gap between the language model and the ad system.
(2) The order of tuning matters, as the fixed-order strategies outperform the mixed strategies. This suggests that the alignment tasks provide a foundational understanding of the ad system, which better prepares the LLM for the main task.

\subsubsection{Effectiveness of DPO}
After tuning the LLM on the auxiliary and main tasks, we apply Reinforcement Learning from Human Feedback (RLHF) to align the language model with business objectives, using both RRHF \cite{yuan2024rrhf} and DPO \cite{rafailov2024direct} techniques. To evaluate the alignment with business values, we use two metrics: Diversity score and LTRR@K.

The Diversity score is calculated based on TopK List Concentration and TopK List Abundance. Concentration is the mean proportion of the most frequent category across all users, while Abundance refers to the mean number of categories present in the top-K list for all users. The Diversity score combines both metrics, where a higher score indicates a more diverse top-K list with lower Concentration and higher Abundance. LTRR@K is computed by Recall@K based on Learning-to-Rank (LTR) labels provided by other retrieval strategies, with higher LTRR@K indicating better alignment with business objectives.

The RLHF results, shown in Figure~\ref{fig_ab_dpo}, lead to the following conclusions:
(1) RLHF improves the alignment of the language model with business values.
(2) DPO outperforms RRHF, likely due to the constraints imposed by the tuned model during the alignment process.

\begin{table}[t]
\caption{ Comparison of different emb models.}
\label{tab_emb_top}
\scalebox{0.95}{\begin{tabular}{c|cccc}
\toprule
\multirow{2}{*}{Emb   model} & \multicolumn{4}{c}{Accuracy}                                                \\ \cline{2-5} 
                             & Top1             & Top10            & Top50            & Top100           \\ \hline
Hunyuan Embedding\cite{sun2024hunyuanlargeopensourcemoemodel}            & \textbf{98.03\%} & {\ul 96.46\%}    & {\ul 86.61\%}    & {\ul  65.67\%}   \\
E5-large-instruct \cite{wang2022text}           & {\ul 97.21\%}    & \textbf{96.48\%} & \textbf{87.42\%} & \textbf{66.38\%} \\
bge-m3\cite{bge-m3}                             & 95.00\%          & 91.07\%          & 78.76\%          & 59.64\%          \\
Bert-Chinese\cite{devlin2018bert}               & {\ul 97.24\%}    & {\ul 92.65\%}    & 74.12\%          & 50.20\%          \\
Electra\_Chinese\cite{clark2020electra}         & 94.60\%          & 85.98\%          & 63.09\%          & 43.62\%          \\
Sentence-Bert/L12\cite{reimers2019sentence}     & 95.40\%          & 90.11\%          & 70.15\%          & 48.44\%          \\
Sentence-Bert/L6\cite{reimers2019sentence}      & 81.97\%          & 72.65\%          & 53.61\%          & 37.97\%          \\
XLNet Chinese\cite{yang2019xlnet}               & 80.31\%          & 63.32\%          & 38.27\%          & 26.14\%          \\
CLIP-32\cite{radford2021learning}               & 80.32\%          & 67.91\%          & 45.90\%          & 29.87\%          \\
CLIP-16\cite{radford2021learning}               & 81.57\%          & 67.40\%          & 45.20\%          & 30.33\%          \\
T5\cite{2020t5}                                 & 67.10\%          & 51.22\%          & 33.90\%          & 23.89\%          \\ \bottomrule
\end{tabular}}
\end{table}

\begin{table}[t]
\caption{Performance comparison of different emb models. }
\label{tab_emb_llm}
\scalebox{0.99}{
\begin{tabular}{c|cccccc}
\toprule
S-IDs & HR@4 & HR@8 & LTRR@4 & LTRR@8 \\ \hline
Hunyuan S-IDs   & 0.1070        & 0.2140        & 0.0294            & 0.0443            \\
E5 S-IDs   & 0.1052        & 0.2140        & 0.0256            & 0.0407            \\ \bottomrule
\end{tabular}
}
\end{table}

\subsubsection{Effectiveness of Feature Embedding}
To assess the effectiveness of the ads' feature embedding used in ads indexing, we compare the performance of different embedding models (emb. models). We first sampled 100 ads from each of the 48 ad categories and retrieved the Top-K nearest ads for each sampled ad based on embedding similarity. The accuracy is defined as the proportion of Top-K retrieved ads that belong to the same category as the original ad. The performance comparison is presented in Table~\ref{tab_emb_top}. Our observations indicate that the E5 model demonstrates the best performance on this task, while the Hunyuan embedding achieves competitive results.

Subsequently, we trained the VQ-VAE using both the E5 and Hunyuan embeddings to infer the ads' S-IDs. We then constructed a trie tree and fine-tuned the LLM using these S-IDs. The performance of the tuned LLM is shown in Table~\ref{tab_emb_llm}. The results indicate that Hunyuan S-IDs exhibit competitive performance in terms of HR@K and outperform in LTRR@K, attributed to their enhanced capability in understanding the ad system.

\begin{table}[t]
\caption{Performance comparison of different codebook size and quantity. }
\label{tab_rqvae_llm}
\scalebox{0.90}{
\begin{tabular}{cc|ccc}
\toprule
\multicolumn{1}{l}{Quantity} & Size & Collision Rate$\downarrow$ & \begin{tabular}[c]{@{}c@{}}Max collision \\ Number$\downarrow$\end{tabular} & \begin{tabular}[c]{@{}c@{}}Average \\ Usage Rate$\uparrow$\end{tabular} \\ \hline
\multirow{3}{*}{3} & 1024 & 4.79\% & 44 & 1.0 \\
 & 512 & 7.95\% & 46 & 1.0 \\
 & 256 & 16.38\% & 67 & 1.0 \\ \hline
\multirow{2}{*}{4} & 1024 & 1.74\% & 25 & 1.0 \\
 & 256 & 3.34\% & 25 & 1.0 \\\bottomrule
\end{tabular}
}
\end{table}

\begin{table}[t]
\caption{Performance comparison of different S-Ids variants. }
\label{tab_sid_llm}
\scalebox{0.85}{
\begin{tabular}{c|ccccc}
\toprule
Variants & HR@1 & HR@4 & HR@8 & NDCG@4 & NDCG@8 \\ \hline
S-IDs & \textbf{0.0800} & \textbf{0.1526} & \textbf{0.1975} & \textbf{0.1194} & \textbf{0.1353} \\ \hline
\multirow{2}{*}{Original Text} & 0.0306 & 0.0551 & 0.0679 & 0.0439 & 0.0487 \\
 & -61.75\% & -63.89\% & -65.62\% & -63.23\% & -64.01\% \\ \hline
\multirow{2}{*}{Compressed Text} & 0.0572 & 0.1102 & 0.1317 & 0.0861 & 0.0967 \\
 & -28.50\% & -27.79\% & -33.32\% & -27.89\% & -28.53\% \\ \bottomrule
\end{tabular}
}
\end{table}


\subsubsection{Sensitivity and Effectiveness of S-IDs}

To evaluate the sensitivity of the ads' Semantic IDs (S-IDs), we varied the codebook size and quantity, assessing the trained ads indexing using three metrics: Collision Rate$\downarrow$, Max Collision Number$\downarrow$, and Average Codebook Usage Rate$\uparrow$. As shown in Table~\ref{tab_rqvae_llm}, increasing the codebook size and quantity reduces ad collisions, indicating that different ads are assigned distinct S-IDs. Notably, the codebook usage rate remains at 100\%, signifying no occurrence of "codebook collapse."

Additionally, to assess the effectiveness of the S-IDs, we removed them and generated both the original and compressed text of ads using the LLM. The performance results are presented in Table~\ref{tab_sid_llm}. The findings indicate a significant performance drop when generating either the original or compressed text, highlighting the importance of S-IDs. Furthermore, the compressed text outperformed better than the original, suggesting that the original text contains redundancy and noise.


\begin{figure}[t]
\centerline{\includegraphics[scale=0.75]{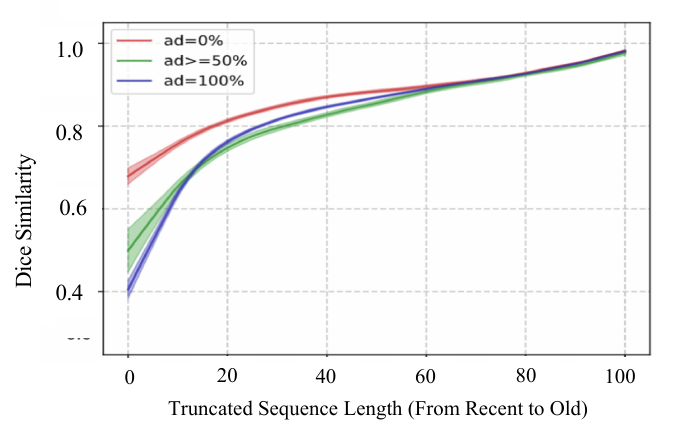}}
\caption{ LLM User Action Response w.r.t Sequence Length.}
\label{fig_ab_user_study}
\end{figure}

\subsubsection{Effectiveness of User Behavior}
To assess how user behavior contributes to LEADRE's ad generation process, we conducted a user study to gain a deeper understanding of the generation mechanism.

To investigate the impact of different positions within user behavior sequences, we conducted the following experiment: For a user $u$ with a behavior sequence $\mathcal{S}_{u,1} = [i^1_u, i^2_u, \dots, i^L_u]$, we sequentially truncated the sequence by removing earlier interactions, forming sequences such as $\mathcal{S}_{u,2} = [i^2_u, i^3_u, \dots, i^L_u]$ and $\mathcal{S}_{u,3} = [i^3_u, i^4_u, \dots, i^L_u]$. LEADRE then made predictions using each truncated sequence, generating top-K lists denoted by $\mathcal{R}_{u,1}$, $\mathcal{R}_{u,2}$, and so on. To quantify the similarity between these lists, we used the Dice similarity coefficient:
\begin{equation}
    \text{Dice}(\mathcal{R}_{u,1}, \mathcal{R}_{u,l}) = \frac{2|\mathcal{R}_{u,1} \cap \mathcal{R}_{u,l}|}{|\mathcal{R}_{u,1}| + |\mathcal{R}_{u,l}|}
\end{equation}
A higher Dice similarity score indicates that the retrieved list generated from the truncated sequence is more similar to that of the complete sequence. This, in turn, suggests that the removed interaction is of lesser importance in influencing the retrieval outcome. We computed the Dice similarity for all users, grouped by the partitioning of their ad behaviors, and plotted the results as a function of truncated sequence length (Figure~\ref{fig_ab_user_study}). The following key observations were made:
(1) Longer behavior sequences capture more user preference information.
(2) LEADRE places more emphasis on the earlier parts of the sequence, particularly the first 20 behaviors, and less emphasis on more recent behaviors.
(3) Ad domain behaviors have a stronger influence on the generated outcomes compared to content domain behaviors, and they encourage LEADRE to focus more on the beginning of the sequence.

\subsubsection{Summary}
In this section, we evaluated the impact of various components, including prompt engineering, tuning tasks, and S-IDs. Our findings are summarized as follows: (1) The S-IDs and the main tuning task are fundamental for ad generation; removing either results in the largest performance drops. (2) Auxiliary tasks, such as explicit and implicit alignment tasks and Direct Preference Optimization (DPO), contribute less than the S-IDs and main task, but still provide a crucial 9\% performance increase. (3) User behaviors from the content and advertising domains are the most significant contributors to ad generation compared to other prompt components.

\subsection{Online Performance}

We conducted a 20\% traffic A/B test on both the Tencent WeChat Channels and Tencent WeChat Moments display advertising systems to evaluate the effectiveness of LEADRE. LEADRE acts as a complementary retrieval sub-brunch. Over several weeks' A/B test on WeChat Channels, LEADRE demonstrated a 1.57\% increase in Gross Merchandise Value (GMV) for serviced users. Similarly, several weeks' A/B test on WeChat Moments showed a 1.17\% increase in GMV for serviced users. These results highlight LEADRE's capability to generate accurate, diverse, and high-value ads that contribute positively to business outcomes. 

LEADRE operates as a retrieval sub-branch, collaborating with other LTR branches to complete the retrieval process. To highlight the competition between branches, we introduce the consumption. Consumption is the proportion of exposed ads exclusively retrieved by a specific branch relative to all exposed ads.
Online A/B test shows that LEADRE achieves a consumption ratio of 7.33\% (Wechat Channels). This demonstrates its ability to retrieve ads missed by other branches, directly contributing to the increase in GMV.

To effectively leverage the capability of LLMs in breaking the "information cocoon" effect, the retrieved ads are further incorporated into the ranking phase as new features. Specifically, on the user side, the retrieved ads are used as additional features to enhance user profiling, while on the item side, the match scores between the retrieved ads and the target ad are introduced as new item-level features. These new features contribute to an extra 1.43\% improvement in GMV on Tencent WeChat Channels.

Currently, LEADRE has been deployed as a complementary retrieval sub-brunch on both Tencent WeChat Channels and Moments, serving billions of users and processing tens of billions of requests each day, highlighting its scalability and practical impact within a large-scale advertising system.

\section{Conclusion}
In this work, we present the first industrial application of generative retrieval in display advertising. To address the challenges of "\textit{How to capture user commercial intent}", "\textit{How to bridge the gap between LLMs and ad-Specific knowledge}", and "\textit{How to efficiently deploy LLMs}", we introduce a novel LLM-based framework called Multi-Faceted Knowledge Enhanced \textbf{L}LM \textbf{E}mpowered Display \textbf{AD}vertisement \textbf{RE}commender system (LEADRE). LEADRE comprises three core modules: the Intent-Aware Prompt Engineering, the Advertising-Specific Knowledge Alignment, and the Latency-Aware Model Deployment.
To evaluate the effectiveness of LEADRE, we conducted extensive experiments, including both offline and online evaluations. The results demonstrate its ability to generate more accurate and diverse ads. Moreover, online A/B test revealed that LEADRE resulted in a 1.57\% increase on Tencent WeChat Channels, as well as a 1.17\% increase on Tencent WeChat Moments in GMV for serviced users.

Looking ahead, our future work will explore the following directions:
(1) Next-N Generation for LLMs: Next-N ad prediction provides a deeper understanding of user intent and offers a more accurate and diverse retrieval list. However, LEADRE currently supports only the generation of a single ad due to its reliance on a statistical trie-tree and fixed tuning tasks. Future efforts will focus on designing appropriate trie-tree structures and constructing tuning tasks for the next-N generation.
(2) Development of More Reasonable S-IDs: Current results indicate that the RQ-VAE tends to allocate most information to the first codebook, while the subsequent codebooks retain limited information. Future research will aim to design an information-controllable quantization model to construct more effective S-IDs.

\section*{Acknowledgments}
This work was supported by National Natural Science Foundation of China (62436010, 62172421) and the Tencent Innovation Fund. Biao Qin is the first corresponding author.

\bibliographystyle{ACM-Reference-Format}
\bibliography{sample}


\begin{thebibliography}{67}


\ifx \showCODEN    \undefined \def \showCODEN     #1{\unskip}     \fi
\ifx \showDOI      \undefined \def \showDOI       #1{#1}\fi
\ifx \showISBNx    \undefined \def \showISBNx     #1{\unskip}     \fi
\ifx \showISBNxiii \undefined \def \showISBNxiii  #1{\unskip}     \fi
\ifx \showISSN     \undefined \def \showISSN      #1{\unskip}     \fi
\ifx \showLCCN     \undefined \def \showLCCN      #1{\unskip}     \fi
\ifx \shownote     \undefined \def \shownote      #1{#1}          \fi
\ifx \showarticletitle \undefined \def \showarticletitle #1{#1}   \fi
\ifx \showURL      \undefined \def \showURL       {\relax}        \fi
\providecommand\bibfield[2]{#2}
\providecommand\bibinfo[2]{#2}
\providecommand\natexlab[1]{#1}
\providecommand\showeprint[2][]{arXiv:#2}

\bibitem[\protect\citeauthoryear{Bao, Zhang, Lin, Zhang, Wang, and Feng}{Bao et~al\mbox{.}}{2024}]%
        {bao2024large}
\bibfield{author}{\bibinfo{person}{Keqin Bao}, \bibinfo{person}{Jizhi Zhang}, \bibinfo{person}{Xinyu Lin}, \bibinfo{person}{Yang Zhang}, \bibinfo{person}{Wenjie Wang}, {and} \bibinfo{person}{Fuli Feng}.} \bibinfo{year}{2024}\natexlab{}.
\newblock \showarticletitle{Large Language Models for Recommendation: Past, Present, and Future}. In \bibinfo{booktitle}{\emph{Proceedings of the 47th International ACM SIGIR Conference on Research and Development in Information Retrieval}}. \bibinfo{pages}{2993--2996}.
\newblock


\bibitem[\protect\citeauthoryear{Bao, Zhang, Zhang, Wang, Feng, and He}{Bao et~al\mbox{.}}{2023}]%
        {bao2023tallrec}
\bibfield{author}{\bibinfo{person}{Keqin Bao}, \bibinfo{person}{Jizhi Zhang}, \bibinfo{person}{Yang Zhang}, \bibinfo{person}{Wenjie Wang}, \bibinfo{person}{Fuli Feng}, {and} \bibinfo{person}{Xiangnan He}.} \bibinfo{year}{2023}\natexlab{}.
\newblock \showarticletitle{Tallrec: An effective and efficient tuning framework to align large language model with recommendation}. In \bibinfo{booktitle}{\emph{Proceedings of the 17th ACM Conference on Recommender Systems}}. \bibinfo{pages}{1007--1014}.
\newblock


\bibitem[\protect\citeauthoryear{Cen, Zhang, Zou, Zhou, Yang, and Tang}{Cen et~al\mbox{.}}{2020}]%
        {cen2020controllable}
\bibfield{author}{\bibinfo{person}{Yukuo Cen}, \bibinfo{person}{Jianwei Zhang}, \bibinfo{person}{Xu Zou}, \bibinfo{person}{Chang Zhou}, \bibinfo{person}{Hongxia Yang}, {and} \bibinfo{person}{Jie Tang}.} \bibinfo{year}{2020}\natexlab{}.
\newblock \showarticletitle{Controllable multi-interest framework for recommendation}. In \bibinfo{booktitle}{\emph{Proceedings of the 26th ACM SIGKDD International Conference on Knowledge Discovery \& Data Mining}}. \bibinfo{pages}{2942--2951}.
\newblock


\bibitem[\protect\citeauthoryear{Chang, Gao, Zheng, Hui, Niu, Song, Jin, and Li}{Chang et~al\mbox{.}}{2021}]%
        {chang2021sequential}
\bibfield{author}{\bibinfo{person}{Jianxin Chang}, \bibinfo{person}{Chen Gao}, \bibinfo{person}{Yu Zheng}, \bibinfo{person}{Yiqun Hui}, \bibinfo{person}{Yanan Niu}, \bibinfo{person}{Yang Song}, \bibinfo{person}{Depeng Jin}, {and} \bibinfo{person}{Yong Li}.} \bibinfo{year}{2021}\natexlab{}.
\newblock \showarticletitle{Sequential recommendation with graph neural networks}. In \bibinfo{booktitle}{\emph{Proceedings of the 44th international ACM SIGIR conference on research and development in information retrieval}}. \bibinfo{pages}{378--387}.
\newblock


\bibitem[\protect\citeauthoryear{Chen, Chen, Zhang, and Tan}{Chen et~al\mbox{.}}{2022}]%
        {chen2022unified}
\bibfield{author}{\bibinfo{person}{Jun Chen}, \bibinfo{person}{Cheng Chen}, \bibinfo{person}{Huayue Zhang}, {and} \bibinfo{person}{Qing Tan}.} \bibinfo{year}{2022}\natexlab{}.
\newblock \showarticletitle{A Unified Framework for Campaign Performance Forecasting in Online Display Advertising}.
\newblock \bibinfo{journal}{\emph{arXiv preprint arXiv:2202.11877}} (\bibinfo{year}{2022}).
\newblock


\bibitem[\protect\citeauthoryear{Chen, Xiao, Zhang, Luo, Lian, and Liu}{Chen et~al\mbox{.}}{2024}]%
        {bge-m3}
\bibfield{author}{\bibinfo{person}{Jianlv Chen}, \bibinfo{person}{Shitao Xiao}, \bibinfo{person}{Peitian Zhang}, \bibinfo{person}{Kun Luo}, \bibinfo{person}{Defu Lian}, {and} \bibinfo{person}{Zheng Liu}.} \bibinfo{year}{2024}\natexlab{}.
\newblock \bibinfo{title}{BGE M3-Embedding: Multi-Lingual, Multi-Functionality, Multi-Granularity Text Embeddings Through Self-Knowledge Distillation}.
\newblock
\newblock
\showeprint[arxiv]{2402.03216}


\bibitem[\protect\citeauthoryear{Clark}{Clark}{2020}]%
        {clark2020electra}
\bibfield{author}{\bibinfo{person}{K Clark}.} \bibinfo{year}{2020}\natexlab{}.
\newblock \showarticletitle{Electra: Pre-training text encoders as discriminators rather than generators}.
\newblock \bibinfo{journal}{\emph{arXiv preprint arXiv:2003.10555}} (\bibinfo{year}{2020}).
\newblock


\bibitem[\protect\citeauthoryear{Cui, Xu, Fei, Cai, Cao, Zhang, and Chen}{Cui et~al\mbox{.}}{2020}]%
        {cui2020personalized}
\bibfield{author}{\bibinfo{person}{Zhihua Cui}, \bibinfo{person}{Xianghua Xu}, \bibinfo{person}{XUE Fei}, \bibinfo{person}{Xingjuan Cai}, \bibinfo{person}{Yang Cao}, \bibinfo{person}{Wensheng Zhang}, {and} \bibinfo{person}{Jinjun Chen}.} \bibinfo{year}{2020}\natexlab{}.
\newblock \showarticletitle{Personalized recommendation system based on collaborative filtering for IoT scenarios}.
\newblock \bibinfo{journal}{\emph{IEEE Transactions on Services Computing}} \bibinfo{volume}{13}, \bibinfo{number}{4} (\bibinfo{year}{2020}), \bibinfo{pages}{685--695}.
\newblock


\bibitem[\protect\citeauthoryear{Devlin}{Devlin}{2018}]%
        {devlin2018bert}
\bibfield{author}{\bibinfo{person}{Jacob Devlin}.} \bibinfo{year}{2018}\natexlab{}.
\newblock \showarticletitle{Bert: Pre-training of deep bidirectional transformers for language understanding}.
\newblock \bibinfo{journal}{\emph{arXiv preprint arXiv:1810.04805}} (\bibinfo{year}{2018}).
\newblock


\bibitem[\protect\citeauthoryear{Dong, Yuan, Yao, Xu, and Zhu}{Dong et~al\mbox{.}}{2020}]%
        {dong2020mamo}
\bibfield{author}{\bibinfo{person}{Manqing Dong}, \bibinfo{person}{Feng Yuan}, \bibinfo{person}{Lina Yao}, \bibinfo{person}{Xiwei Xu}, {and} \bibinfo{person}{Liming Zhu}.} \bibinfo{year}{2020}\natexlab{}.
\newblock \showarticletitle{Mamo: Memory-augmented meta-optimization for cold-start recommendation}. In \bibinfo{booktitle}{\emph{Proceedings of the 26th ACM SIGKDD international conference on knowledge discovery \& data mining}}. \bibinfo{pages}{688--697}.
\newblock


\bibitem[\protect\citeauthoryear{Geng, Liu, Fu, Ge, and Zhang}{Geng et~al\mbox{.}}{2022}]%
        {geng2022recommendation}
\bibfield{author}{\bibinfo{person}{Shijie Geng}, \bibinfo{person}{Shuchang Liu}, \bibinfo{person}{Zuohui Fu}, \bibinfo{person}{Yingqiang Ge}, {and} \bibinfo{person}{Yongfeng Zhang}.} \bibinfo{year}{2022}\natexlab{}.
\newblock \showarticletitle{Recommendation as language processing (rlp): A unified pretrain, personalized prompt \& predict paradigm (p5)}. In \bibinfo{booktitle}{\emph{Proceedings of the 16th ACM Conference on Recommender Systems}}. \bibinfo{pages}{299--315}.
\newblock


\bibitem[\protect\citeauthoryear{Gharibshah, Zhu, Hainline, and Conway}{Gharibshah et~al\mbox{.}}{2020}]%
        {gharibshah2020deep}
\bibfield{author}{\bibinfo{person}{Zhabiz Gharibshah}, \bibinfo{person}{Xingquan Zhu}, \bibinfo{person}{Arthur Hainline}, {and} \bibinfo{person}{Michael Conway}.} \bibinfo{year}{2020}\natexlab{}.
\newblock \showarticletitle{Deep learning for user interest and response prediction in online display advertising}.
\newblock \bibinfo{journal}{\emph{Data Science and Engineering}} \bibinfo{volume}{5}, \bibinfo{number}{1} (\bibinfo{year}{2020}), \bibinfo{pages}{12--26}.
\newblock


\bibitem[\protect\citeauthoryear{Guo, Xu, Duan, Yin, and McAuley}{Guo et~al\mbox{.}}{2023}]%
        {guo2023longcoder}
\bibfield{author}{\bibinfo{person}{Daya Guo}, \bibinfo{person}{Canwen Xu}, \bibinfo{person}{Nan Duan}, \bibinfo{person}{Jian Yin}, {and} \bibinfo{person}{Julian McAuley}.} \bibinfo{year}{2023}\natexlab{}.
\newblock \showarticletitle{Longcoder: A long-range pre-trained language model for code completion}. In \bibinfo{booktitle}{\emph{International Conference on Machine Learning}}. PMLR, \bibinfo{pages}{12098--12107}.
\newblock


\bibitem[\protect\citeauthoryear{Harte, Zorgdrager, Louridas, Katsifodimos, Jannach, and Fragkoulis}{Harte et~al\mbox{.}}{2023}]%
        {harte2023leveraging}
\bibfield{author}{\bibinfo{person}{Jesse Harte}, \bibinfo{person}{Wouter Zorgdrager}, \bibinfo{person}{Panos Louridas}, \bibinfo{person}{Asterios Katsifodimos}, \bibinfo{person}{Dietmar Jannach}, {and} \bibinfo{person}{Marios Fragkoulis}.} \bibinfo{year}{2023}\natexlab{}.
\newblock \showarticletitle{Leveraging large language models for sequential recommendation}. In \bibinfo{booktitle}{\emph{Proceedings of the 17th ACM Conference on Recommender Systems}}. \bibinfo{pages}{1096--1102}.
\newblock


\bibitem[\protect\citeauthoryear{He, Xie, Jha, Steck, Liang, Feng, Majumder, Kallus, and McAuley}{He et~al\mbox{.}}{2023}]%
        {he2023large}
\bibfield{author}{\bibinfo{person}{Zhankui He}, \bibinfo{person}{Zhouhang Xie}, \bibinfo{person}{Rahul Jha}, \bibinfo{person}{Harald Steck}, \bibinfo{person}{Dawen Liang}, \bibinfo{person}{Yesu Feng}, \bibinfo{person}{Bodhisattwa~Prasad Majumder}, \bibinfo{person}{Nathan Kallus}, {and} \bibinfo{person}{Julian McAuley}.} \bibinfo{year}{2023}\natexlab{}.
\newblock \showarticletitle{Large language models as zero-shot conversational recommenders}. In \bibinfo{booktitle}{\emph{Proceedings of the 32nd ACM international conference on information and knowledge management}}.
\newblock


\bibitem[\protect\citeauthoryear{Hokamp and Liu}{Hokamp and Liu}{2017}]%
        {hokamp2017lexically}
\bibfield{author}{\bibinfo{person}{Chris Hokamp} {and} \bibinfo{person}{Qun Liu}.} \bibinfo{year}{2017}\natexlab{}.
\newblock \showarticletitle{Lexically constrained decoding for sequence generation using grid beam search}.
\newblock \bibinfo{journal}{\emph{arXiv preprint arXiv:1704.07138}} (\bibinfo{year}{2017}).
\newblock


\bibitem[\protect\citeauthoryear{Hu, Khayrallah, Culkin, Xia, Chen, Post, and Van~Durme}{Hu et~al\mbox{.}}{2019}]%
        {hu2019improved}
\bibfield{author}{\bibinfo{person}{J~Edward Hu}, \bibinfo{person}{Huda Khayrallah}, \bibinfo{person}{Ryan Culkin}, \bibinfo{person}{Patrick Xia}, \bibinfo{person}{Tongfei Chen}, \bibinfo{person}{Matt Post}, {and} \bibinfo{person}{Benjamin Van~Durme}.} \bibinfo{year}{2019}\natexlab{}.
\newblock \showarticletitle{Improved lexically constrained decoding for translation and monolingual rewriting}. In \bibinfo{booktitle}{\emph{Proceedings of the 2019 Conference of the North American Chapter of the Association for Computational Linguistics: Human Language Technologies, Volume 1 (Long and Short Papers)}}. \bibinfo{pages}{839--850}.
\newblock


\bibitem[\protect\citeauthoryear{Huang, Gu, Hou, Wu, Wang, Yu, and Han}{Huang et~al\mbox{.}}{2022}]%
        {huang2022large}
\bibfield{author}{\bibinfo{person}{Jiaxin Huang}, \bibinfo{person}{Shixiang~Shane Gu}, \bibinfo{person}{Le Hou}, \bibinfo{person}{Yuexin Wu}, \bibinfo{person}{Xuezhi Wang}, \bibinfo{person}{Hongkun Yu}, {and} \bibinfo{person}{Jiawei Han}.} \bibinfo{year}{2022}\natexlab{}.
\newblock \showarticletitle{Large language models can self-improve}.
\newblock \bibinfo{journal}{\emph{arXiv preprint arXiv:2210.11610}} (\bibinfo{year}{2022}).
\newblock


\bibitem[\protect\citeauthoryear{Huang, Sharma, Sun, Xia, Zhang, Pronin, Padmanabhan, Ottaviano, and Yang}{Huang et~al\mbox{.}}{2020}]%
        {huang2020embedding}
\bibfield{author}{\bibinfo{person}{Jui-Ting Huang}, \bibinfo{person}{Ashish Sharma}, \bibinfo{person}{Shuying Sun}, \bibinfo{person}{Li Xia}, \bibinfo{person}{David Zhang}, \bibinfo{person}{Philip Pronin}, \bibinfo{person}{Janani Padmanabhan}, \bibinfo{person}{Giuseppe Ottaviano}, {and} \bibinfo{person}{Linjun Yang}.} \bibinfo{year}{2020}\natexlab{}.
\newblock \showarticletitle{Embedding-based retrieval in facebook search}. In \bibinfo{booktitle}{\emph{Proceedings of the 26th ACM SIGKDD International Conference on Knowledge Discovery \& Data Mining}}. \bibinfo{pages}{2553--2561}.
\newblock


\bibitem[\protect\citeauthoryear{Ji, Tang, Chen, Deng, Hu, and Su}{Ji et~al\mbox{.}}{2024}]%
        {ji2024neural}
\bibfield{author}{\bibinfo{person}{Houye Ji}, \bibinfo{person}{Ye Tang}, \bibinfo{person}{Zhaoxin Chen}, \bibinfo{person}{Lixi Deng}, \bibinfo{person}{Jun Hu}, {and} \bibinfo{person}{Lei Su}.} \bibinfo{year}{2024}\natexlab{}.
\newblock \showarticletitle{Neural Graph Matching for Video Retrieval in Large-Scale Video-driven E-commerce}.
\newblock \bibinfo{journal}{\emph{arXiv preprint arXiv:2408.00346}} (\bibinfo{year}{2024}).
\newblock


\bibitem[\protect\citeauthoryear{Kang and McAuley}{Kang and McAuley}{2018}]%
        {kang2018self}
\bibfield{author}{\bibinfo{person}{Wang-Cheng Kang} {and} \bibinfo{person}{Julian McAuley}.} \bibinfo{year}{2018}\natexlab{}.
\newblock \showarticletitle{Self-attentive sequential recommendation}. In \bibinfo{booktitle}{\emph{2018 IEEE international conference on data mining (ICDM)}}. IEEE, \bibinfo{pages}{197--206}.
\newblock


\bibitem[\protect\citeauthoryear{Koh, Ju, Liu, and Pan}{Koh et~al\mbox{.}}{2022}]%
        {koh2022empirical}
\bibfield{author}{\bibinfo{person}{Huan~Yee Koh}, \bibinfo{person}{Jiaxin Ju}, \bibinfo{person}{Ming Liu}, {and} \bibinfo{person}{Shirui Pan}.} \bibinfo{year}{2022}\natexlab{}.
\newblock \showarticletitle{An empirical survey on long document summarization: Datasets, models, and metrics}.
\newblock \bibinfo{journal}{\emph{ACM computing surveys}} \bibinfo{volume}{55}, \bibinfo{number}{8} (\bibinfo{year}{2022}), \bibinfo{pages}{1--35}.
\newblock


\bibitem[\protect\citeauthoryear{Kumar, Zhang, and Leskovec}{Kumar et~al\mbox{.}}{2019}]%
        {kumar2019predicting}
\bibfield{author}{\bibinfo{person}{Srijan Kumar}, \bibinfo{person}{Xikun Zhang}, {and} \bibinfo{person}{Jure Leskovec}.} \bibinfo{year}{2019}\natexlab{}.
\newblock \showarticletitle{Predicting dynamic embedding trajectory in temporal interaction networks}. In \bibinfo{booktitle}{\emph{Proceedings of the 25th ACM SIGKDD international conference on knowledge discovery \& data mining}}. \bibinfo{pages}{1269--1278}.
\newblock


\bibitem[\protect\citeauthoryear{Lee, Kim, Kim, Cho, and Han}{Lee et~al\mbox{.}}{2022}]%
        {lee2022autoregressive}
\bibfield{author}{\bibinfo{person}{Doyup Lee}, \bibinfo{person}{Chiheon Kim}, \bibinfo{person}{Saehoon Kim}, \bibinfo{person}{Minsu Cho}, {and} \bibinfo{person}{Wook-Shin Han}.} \bibinfo{year}{2022}\natexlab{}.
\newblock \showarticletitle{Autoregressive image generation using residual quantization}. In \bibinfo{booktitle}{\emph{Proceedings of the IEEE/CVF Conference on Computer Vision and Pattern Recognition}}.
\newblock


\bibitem[\protect\citeauthoryear{Li, Liu, Li, Xu, Cao, Li, Jiang, Wang, Zhu, Gai, et~al\mbox{.}}{Li et~al\mbox{.}}{2021}]%
        {li2021truncation}
\bibfield{author}{\bibinfo{person}{Jin Li}, \bibinfo{person}{Jie Liu}, \bibinfo{person}{Shangzhou Li}, \bibinfo{person}{Yao Xu}, \bibinfo{person}{Ran Cao}, \bibinfo{person}{Qi Li}, \bibinfo{person}{Biye Jiang}, \bibinfo{person}{Guan Wang}, \bibinfo{person}{Han Zhu}, \bibinfo{person}{Kun Gai}, {et~al\mbox{.}}} \bibinfo{year}{2021}\natexlab{}.
\newblock \showarticletitle{Truncation-Free Matching System for Display Advertising at Alibaba}.
\newblock \bibinfo{journal}{\emph{arXiv preprint arXiv:2102.09283}} (\bibinfo{year}{2021}).
\newblock


\bibitem[\protect\citeauthoryear{Li, Wang, Li, Fu, Shen, Shang, and McAuley}{Li et~al\mbox{.}}{2023}]%
        {li2023text}
\bibfield{author}{\bibinfo{person}{Jiacheng Li}, \bibinfo{person}{Ming Wang}, \bibinfo{person}{Jin Li}, \bibinfo{person}{Jinmiao Fu}, \bibinfo{person}{Xin Shen}, \bibinfo{person}{Jingbo Shang}, {and} \bibinfo{person}{Julian McAuley}.} \bibinfo{year}{2023}\natexlab{}.
\newblock \showarticletitle{Text is all you need: Learning language representations for sequential recommendation}. In \bibinfo{booktitle}{\emph{Proceedings of the 29th ACM SIGKDD Conference on Knowledge Discovery and Data Mining}}. \bibinfo{pages}{1258--1267}.
\newblock


\bibitem[\protect\citeauthoryear{Li and Zhang}{Li and Zhang}{2024}]%
        {li2024planning}
\bibfield{author}{\bibinfo{person}{Kunze Li} {and} \bibinfo{person}{Yu Zhang}.} \bibinfo{year}{2024}\natexlab{}.
\newblock \showarticletitle{Planning First, Question Second: An LLM-Guided Method for Controllable Question Generation}. In \bibinfo{booktitle}{\emph{Findings of the Association for Computational Linguistics ACL 2024}}. \bibinfo{pages}{4715--4729}.
\newblock


\bibitem[\protect\citeauthoryear{Lin, Wang, Li, Feng, Ng, and Chua}{Lin et~al\mbox{.}}{2024}]%
        {lin2024bridging}
\bibfield{author}{\bibinfo{person}{Xinyu Lin}, \bibinfo{person}{Wenjie Wang}, \bibinfo{person}{Yongqi Li}, \bibinfo{person}{Fuli Feng}, \bibinfo{person}{See-Kiong Ng}, {and} \bibinfo{person}{Tat-Seng Chua}.} \bibinfo{year}{2024}\natexlab{}.
\newblock \showarticletitle{Bridging items and language: A transition paradigm for large language model-based recommendation}. In \bibinfo{booktitle}{\emph{Proceedings of the 30th ACM SIGKDD Conference on Knowledge Discovery and Data Mining}}. \bibinfo{pages}{1816--1826}.
\newblock


\bibitem[\protect\citeauthoryear{Liu, Li, Cai, Dong, Zhu, and Shang}{Liu et~al\mbox{.}}{2021}]%
        {liu2021noninvasive}
\bibfield{author}{\bibinfo{person}{Chang Liu}, \bibinfo{person}{Xiaoguang Li}, \bibinfo{person}{Guohao Cai}, \bibinfo{person}{Zhenhua Dong}, \bibinfo{person}{Hong Zhu}, {and} \bibinfo{person}{Lifeng Shang}.} \bibinfo{year}{2021}\natexlab{}.
\newblock \showarticletitle{Noninvasive self-attention for side information fusion in sequential recommendation}. In \bibinfo{booktitle}{\emph{Proceedings of the AAAI conference on artificial intelligence}}, Vol.~\bibinfo{volume}{35}. \bibinfo{pages}{4249--4256}.
\newblock


\bibitem[\protect\citeauthoryear{Liu, Wu, Wang, Li, and Wang}{Liu et~al\mbox{.}}{2016}]%
        {liu2016context}
\bibfield{author}{\bibinfo{person}{Qiang Liu}, \bibinfo{person}{Shu Wu}, \bibinfo{person}{Diyi Wang}, \bibinfo{person}{Zhaokang Li}, {and} \bibinfo{person}{Liang Wang}.} \bibinfo{year}{2016}\natexlab{}.
\newblock \showarticletitle{Context-aware sequential recommendation}. In \bibinfo{booktitle}{\emph{2016 IEEE 16th International Conference on Data Mining (ICDM)}}. IEEE, \bibinfo{pages}{1053--1058}.
\newblock


\bibitem[\protect\citeauthoryear{Nam, Macvean, Hellendoorn, Vasilescu, and Myers}{Nam et~al\mbox{.}}{2024}]%
        {nam2024using}
\bibfield{author}{\bibinfo{person}{Daye Nam}, \bibinfo{person}{Andrew Macvean}, \bibinfo{person}{Vincent Hellendoorn}, \bibinfo{person}{Bogdan Vasilescu}, {and} \bibinfo{person}{Brad Myers}.} \bibinfo{year}{2024}\natexlab{}.
\newblock \showarticletitle{Using an llm to help with code understanding}. In \bibinfo{booktitle}{\emph{Proceedings of the IEEE/ACM 46th International Conference on Software Engineering}}. \bibinfo{pages}{1--13}.
\newblock


\bibitem[\protect\citeauthoryear{Post and Vilar}{Post and Vilar}{2018}]%
        {post2018fast}
\bibfield{author}{\bibinfo{person}{Matt Post} {and} \bibinfo{person}{David Vilar}.} \bibinfo{year}{2018}\natexlab{}.
\newblock \showarticletitle{Fast lexically constrained decoding with dynamic beam allocation for neural machine translation}.
\newblock \bibinfo{journal}{\emph{arXiv preprint arXiv:1804.06609}} (\bibinfo{year}{2018}).
\newblock


\bibitem[\protect\citeauthoryear{Qin and Eisner}{Qin and Eisner}{2021}]%
        {qin2021learning}
\bibfield{author}{\bibinfo{person}{Guanghui Qin} {and} \bibinfo{person}{Jason Eisner}.} \bibinfo{year}{2021}\natexlab{}.
\newblock \showarticletitle{Learning how to ask: Querying LMs with mixtures of soft prompts}.
\newblock \bibinfo{journal}{\emph{arXiv preprint arXiv:2104.06599}} (\bibinfo{year}{2021}).
\newblock


\bibitem[\protect\citeauthoryear{Radford, Kim, Hallacy, Ramesh, Goh, Agarwal, Sastry, Askell, Mishkin, Clark, et~al\mbox{.}}{Radford et~al\mbox{.}}{2021}]%
        {radford2021learning}
\bibfield{author}{\bibinfo{person}{Alec Radford}, \bibinfo{person}{Jong~Wook Kim}, \bibinfo{person}{Chris Hallacy}, \bibinfo{person}{Aditya Ramesh}, \bibinfo{person}{Gabriel Goh}, \bibinfo{person}{Sandhini Agarwal}, \bibinfo{person}{Girish Sastry}, \bibinfo{person}{Amanda Askell}, \bibinfo{person}{Pamela Mishkin}, \bibinfo{person}{Jack Clark}, {et~al\mbox{.}}} \bibinfo{year}{2021}\natexlab{}.
\newblock \showarticletitle{Learning transferable visual models from natural language supervision}. In \bibinfo{booktitle}{\emph{International conference on machine learning}}. PMLR, \bibinfo{pages}{8748--8763}.
\newblock


\bibitem[\protect\citeauthoryear{Rafailov, Sharma, Mitchell, Manning, Ermon, and Finn}{Rafailov et~al\mbox{.}}{2024}]%
        {rafailov2024direct}
\bibfield{author}{\bibinfo{person}{Rafael Rafailov}, \bibinfo{person}{Archit Sharma}, \bibinfo{person}{Eric Mitchell}, \bibinfo{person}{Christopher~D Manning}, \bibinfo{person}{Stefano Ermon}, {and} \bibinfo{person}{Chelsea Finn}.} \bibinfo{year}{2024}\natexlab{}.
\newblock \showarticletitle{Direct preference optimization: Your language model is secretly a reward model}.
\newblock \bibinfo{journal}{\emph{Advances in Neural Information Processing Systems}}  \bibinfo{volume}{36} (\bibinfo{year}{2024}).
\newblock


\bibitem[\protect\citeauthoryear{Raffel, Shazeer, Roberts, Lee, Narang, Matena, Zhou, Li, and Liu}{Raffel et~al\mbox{.}}{2020}]%
        {2020t5}
\bibfield{author}{\bibinfo{person}{Colin Raffel}, \bibinfo{person}{Noam Shazeer}, \bibinfo{person}{Adam Roberts}, \bibinfo{person}{Katherine Lee}, \bibinfo{person}{Sharan Narang}, \bibinfo{person}{Michael Matena}, \bibinfo{person}{Yanqi Zhou}, \bibinfo{person}{Wei Li}, {and} \bibinfo{person}{Peter~J. Liu}.} \bibinfo{year}{2020}\natexlab{}.
\newblock \showarticletitle{Exploring the Limits of Transfer Learning with a Unified Text-to-Text Transformer}.
\newblock \bibinfo{journal}{\emph{Journal of Machine Learning Research}} \bibinfo{volume}{21}, \bibinfo{number}{140} (\bibinfo{year}{2020}), \bibinfo{pages}{1--67}.
\newblock
\urldef\tempurl%
\url{http://jmlr.org/papers/v21/20-074.html}
\showURL{%
\tempurl}


\bibitem[\protect\citeauthoryear{Rajput, Mehta, Singh, Hulikal~Keshavan, Vu, Heldt, Hong, Tay, Tran, Samost, et~al\mbox{.}}{Rajput et~al\mbox{.}}{2024}]%
        {rajput2024recommender}
\bibfield{author}{\bibinfo{person}{Shashank Rajput}, \bibinfo{person}{Nikhil Mehta}, \bibinfo{person}{Anima Singh}, \bibinfo{person}{Raghunandan Hulikal~Keshavan}, \bibinfo{person}{Trung Vu}, \bibinfo{person}{Lukasz Heldt}, \bibinfo{person}{Lichan Hong}, \bibinfo{person}{Yi Tay}, \bibinfo{person}{Vinh Tran}, \bibinfo{person}{Jonah Samost}, {et~al\mbox{.}}} \bibinfo{year}{2024}\natexlab{}.
\newblock \showarticletitle{Recommender systems with generative retrieval}.
\newblock \bibinfo{journal}{\emph{Advances in Neural Information Processing Systems}}  \bibinfo{volume}{36} (\bibinfo{year}{2024}).
\newblock


\bibitem[\protect\citeauthoryear{Reimers}{Reimers}{2019}]%
        {reimers2019sentence}
\bibfield{author}{\bibinfo{person}{N Reimers}.} \bibinfo{year}{2019}\natexlab{}.
\newblock \showarticletitle{Sentence-BERT: Sentence Embeddings using Siamese BERT-Networks}.
\newblock \bibinfo{journal}{\emph{arXiv preprint arXiv:1908.10084}} (\bibinfo{year}{2019}).
\newblock


\bibitem[\protect\citeauthoryear{Ren, Wei, Xia, Su, Cheng, Wang, Yin, and Huang}{Ren et~al\mbox{.}}{2024}]%
        {ren2024representation}
\bibfield{author}{\bibinfo{person}{Xubin Ren}, \bibinfo{person}{Wei Wei}, \bibinfo{person}{Lianghao Xia}, \bibinfo{person}{Lixin Su}, \bibinfo{person}{Suqi Cheng}, \bibinfo{person}{Junfeng Wang}, \bibinfo{person}{Dawei Yin}, {and} \bibinfo{person}{Chao Huang}.} \bibinfo{year}{2024}\natexlab{}.
\newblock \showarticletitle{Representation learning with large language models for recommendation}. In \bibinfo{booktitle}{\emph{Proceedings of the ACM on Web Conference 2024}}.
\newblock


\bibitem[\protect\citeauthoryear{Singh, Nanavati, Kar, and Gupta}{Singh et~al\mbox{.}}{2023}]%
        {singh2023maximize}
\bibfield{author}{\bibinfo{person}{Vinay Singh}, \bibinfo{person}{Brijesh Nanavati}, \bibinfo{person}{Arpan~Kumar Kar}, {and} \bibinfo{person}{Agam Gupta}.} \bibinfo{year}{2023}\natexlab{}.
\newblock \showarticletitle{How to maximize clicks for display advertisement in digital marketing? A reinforcement learning approach}.
\newblock \bibinfo{journal}{\emph{Information Systems Frontiers}} \bibinfo{volume}{25}, \bibinfo{number}{4} (\bibinfo{year}{2023}), \bibinfo{pages}{1621--1638}.
\newblock


\bibitem[\protect\citeauthoryear{Song, Yang, Guo, Shen, Jiang, and Wang}{Song et~al\mbox{.}}{2024}]%
        {song2024multi}
\bibfield{author}{\bibinfo{person}{Derun Song}, \bibinfo{person}{Enneng Yang}, \bibinfo{person}{Guibing Guo}, \bibinfo{person}{Li Shen}, \bibinfo{person}{Linying Jiang}, {and} \bibinfo{person}{Xingwei Wang}.} \bibinfo{year}{2024}\natexlab{}.
\newblock \showarticletitle{Multi-scenario and multi-task aware feature interaction for recommendation system}.
\newblock \bibinfo{journal}{\emph{ACM Transactions on Knowledge Discovery from Data}} \bibinfo{volume}{18}, \bibinfo{number}{6} (\bibinfo{year}{2024}), \bibinfo{pages}{1--20}.
\newblock


\bibitem[\protect\citeauthoryear{Sun, Liu, Wu, Pei, Lin, Ou, and Jiang}{Sun et~al\mbox{.}}{2019}]%
        {sun2019bert4rec}
\bibfield{author}{\bibinfo{person}{Fei Sun}, \bibinfo{person}{Jun Liu}, \bibinfo{person}{Jian Wu}, \bibinfo{person}{Changhua Pei}, \bibinfo{person}{Xiao Lin}, \bibinfo{person}{Wenwu Ou}, {and} \bibinfo{person}{Peng Jiang}.} \bibinfo{year}{2019}\natexlab{}.
\newblock \showarticletitle{BERT4Rec: Sequential recommendation with bidirectional encoder representations from transformer}. In \bibinfo{booktitle}{\emph{Proceedings of the 28th ACM international conference on information and knowledge management}}. \bibinfo{pages}{1441--1450}.
\newblock


\bibitem[\protect\citeauthoryear{Sun, Chen, Huang, Xie, Zhu, Zhang, Li, Yang, Han, Shu, et~al\mbox{.}}{Sun et~al\mbox{.}}{2024}]%
        {sun2024hunyuanlargeopensourcemoemodel}
\bibfield{author}{\bibinfo{person}{Xingwu Sun}, \bibinfo{person}{Yanfeng Chen}, \bibinfo{person}{Yiqing Huang}, \bibinfo{person}{Ruobing Xie}, \bibinfo{person}{Jiaqi Zhu}, \bibinfo{person}{Kai Zhang}, \bibinfo{person}{Shuaipeng Li}, \bibinfo{person}{Zhen Yang}, \bibinfo{person}{Jonny Han}, \bibinfo{person}{Xiaobo Shu}, {et~al\mbox{.}}} \bibinfo{year}{2024}\natexlab{}.
\newblock \bibinfo{title}{Hunyuan-Large: An Open-Source MoE Model with 52 Billion Activated Parameters by Tencent}.
\newblock
\newblock
\showeprint[arxiv]{2411.02265}


\bibitem[\protect\citeauthoryear{Vu, Lester, Constant, Al-Rfou, and Cer}{Vu et~al\mbox{.}}{2021}]%
        {vu2021spot}
\bibfield{author}{\bibinfo{person}{Tu Vu}, \bibinfo{person}{Brian Lester}, \bibinfo{person}{Noah Constant}, \bibinfo{person}{Rami Al-Rfou}, {and} \bibinfo{person}{Daniel Cer}.} \bibinfo{year}{2021}\natexlab{}.
\newblock \showarticletitle{Spot: Better frozen model adaptation through soft prompt transfer}.
\newblock \bibinfo{journal}{\emph{arXiv preprint arXiv:2110.07904}} (\bibinfo{year}{2021}).
\newblock


\bibitem[\protect\citeauthoryear{Wang, Li, and Li}{Wang et~al\mbox{.}}{2023}]%
        {wang2023enabling}
\bibfield{author}{\bibinfo{person}{Bryan Wang}, \bibinfo{person}{Gang Li}, {and} \bibinfo{person}{Yang Li}.} \bibinfo{year}{2023}\natexlab{}.
\newblock \showarticletitle{Enabling conversational interaction with mobile ui using large language models}. In \bibinfo{booktitle}{\emph{Proceedings of the 2023 CHI Conference on Human Factors in Computing Systems}}. \bibinfo{pages}{1--17}.
\newblock


\bibitem[\protect\citeauthoryear{Wang, Yang, Huang, Jiao, Yang, Jiang, Majumder, and Wei}{Wang et~al\mbox{.}}{2022}]%
        {wang2022text}
\bibfield{author}{\bibinfo{person}{Liang Wang}, \bibinfo{person}{Nan Yang}, \bibinfo{person}{Xiaolong Huang}, \bibinfo{person}{Binxing Jiao}, \bibinfo{person}{Linjun Yang}, \bibinfo{person}{Daxin Jiang}, \bibinfo{person}{Rangan Majumder}, {and} \bibinfo{person}{Furu Wei}.} \bibinfo{year}{2022}\natexlab{}.
\newblock \showarticletitle{Text embeddings by weakly-supervised contrastive pre-training}.
\newblock \bibinfo{journal}{\emph{arXiv preprint arXiv:2212.03533}} (\bibinfo{year}{2022}).
\newblock


\bibitem[\protect\citeauthoryear{Wei, Ren, Tang, Wang, Su, Cheng, Wang, Yin, and Huang}{Wei et~al\mbox{.}}{2024}]%
        {wei2024llmrec}
\bibfield{author}{\bibinfo{person}{Wei Wei}, \bibinfo{person}{Xubin Ren}, \bibinfo{person}{Jiabin Tang}, \bibinfo{person}{Qinyong Wang}, \bibinfo{person}{Lixin Su}, \bibinfo{person}{Suqi Cheng}, \bibinfo{person}{Junfeng Wang}, \bibinfo{person}{Dawei Yin}, {and} \bibinfo{person}{Chao Huang}.} \bibinfo{year}{2024}\natexlab{}.
\newblock \showarticletitle{Llmrec: Large language models with graph augmentation for recommendation}. In \bibinfo{booktitle}{\emph{Proceedings of the 17th ACM International Conference on Web Search and Data Mining}}. \bibinfo{pages}{806--815}.
\newblock


\bibitem[\protect\citeauthoryear{Wei, Wang, Li, Nie, Li, Li, and Chua}{Wei et~al\mbox{.}}{2021}]%
        {wei2021contrastive}
\bibfield{author}{\bibinfo{person}{Yinwei Wei}, \bibinfo{person}{Xiang Wang}, \bibinfo{person}{Qi Li}, \bibinfo{person}{Liqiang Nie}, \bibinfo{person}{Yan Li}, \bibinfo{person}{Xuanping Li}, {and} \bibinfo{person}{Tat-Seng Chua}.} \bibinfo{year}{2021}\natexlab{}.
\newblock \showarticletitle{Contrastive learning for cold-start recommendation}. In \bibinfo{booktitle}{\emph{Proceedings of the 29th ACM International Conference on Multimedia}}. \bibinfo{pages}{5382--5390}.
\newblock


\bibitem[\protect\citeauthoryear{Wu, Li, Hsieh, and Sharpnack}{Wu et~al\mbox{.}}{2020}]%
        {wu2020sse}
\bibfield{author}{\bibinfo{person}{Liwei Wu}, \bibinfo{person}{Shuqing Li}, \bibinfo{person}{Cho-Jui Hsieh}, {and} \bibinfo{person}{James Sharpnack}.} \bibinfo{year}{2020}\natexlab{}.
\newblock \showarticletitle{SSE-PT: Sequential recommendation via personalized transformer}. In \bibinfo{booktitle}{\emph{Proceedings of the 14th ACM conference on recommender systems}}. \bibinfo{pages}{328--337}.
\newblock


\bibitem[\protect\citeauthoryear{Wu, Zheng, Qiu, Wang, Gu, Shen, Qin, Zhu, Zhu, Liu, et~al\mbox{.}}{Wu et~al\mbox{.}}{2024}]%
        {wu2024survey}
\bibfield{author}{\bibinfo{person}{Likang Wu}, \bibinfo{person}{Zhi Zheng}, \bibinfo{person}{Zhaopeng Qiu}, \bibinfo{person}{Hao Wang}, \bibinfo{person}{Hongchao Gu}, \bibinfo{person}{Tingjia Shen}, \bibinfo{person}{Chuan Qin}, \bibinfo{person}{Chen Zhu}, \bibinfo{person}{Hengshu Zhu}, \bibinfo{person}{Qi Liu}, {et~al\mbox{.}}} \bibinfo{year}{2024}\natexlab{}.
\newblock \showarticletitle{A survey on large language models for recommendation}.
\newblock \bibinfo{journal}{\emph{World Wide Web}} \bibinfo{volume}{27}, \bibinfo{number}{5} (\bibinfo{year}{2024}), \bibinfo{pages}{60}.
\newblock


\bibitem[\protect\citeauthoryear{Xie, Liu, Wang, Liu, Zhang, and Lin}{Xie et~al\mbox{.}}{2022a}]%
        {xie2022contrastive}
\bibfield{author}{\bibinfo{person}{Ruobing Xie}, \bibinfo{person}{Qi Liu}, \bibinfo{person}{Liangdong Wang}, \bibinfo{person}{Shukai Liu}, \bibinfo{person}{Bo Zhang}, {and} \bibinfo{person}{Leyu Lin}.} \bibinfo{year}{2022}\natexlab{a}.
\newblock \showarticletitle{Contrastive cross-domain recommendation in matching}. In \bibinfo{booktitle}{\emph{Proceedings of the 28th ACM SIGKDD conference on knowledge discovery and data mining}}.
\newblock


\bibitem[\protect\citeauthoryear{Xie, Zhou, and Kim}{Xie et~al\mbox{.}}{2022b}]%
        {xie2022decoupled}
\bibfield{author}{\bibinfo{person}{Yueqi Xie}, \bibinfo{person}{Peilin Zhou}, {and} \bibinfo{person}{Sunghun Kim}.} \bibinfo{year}{2022}\natexlab{b}.
\newblock \showarticletitle{Decoupled side information fusion for sequential recommendation}. In \bibinfo{booktitle}{\emph{Proceedings of the 45th international ACM SIGIR conference on research and development in information retrieval}}. \bibinfo{pages}{1611--1621}.
\newblock


\bibitem[\protect\citeauthoryear{Xu, Zhao, Liu, Xu, S.~Sheng, Cui, Zhou, and Xiong}{Xu et~al\mbox{.}}{2019b}]%
        {xu2019recurrent}
\bibfield{author}{\bibinfo{person}{Chengfeng Xu}, \bibinfo{person}{Pengpeng Zhao}, \bibinfo{person}{Yanchi Liu}, \bibinfo{person}{Jiajie Xu}, \bibinfo{person}{Victor S~Sheng S.~Sheng}, \bibinfo{person}{Zhiming Cui}, \bibinfo{person}{Xiaofang Zhou}, {and} \bibinfo{person}{Hui Xiong}.} \bibinfo{year}{2019}\natexlab{b}.
\newblock \showarticletitle{Recurrent convolutional neural network for sequential recommendation}. In \bibinfo{booktitle}{\emph{The world wide web conference}}. \bibinfo{pages}{3398--3404}.
\newblock


\bibitem[\protect\citeauthoryear{Xu, Liu, and Xu}{Xu et~al\mbox{.}}{2019a}]%
        {xu2019survey}
\bibfield{author}{\bibinfo{person}{Mingming Xu}, \bibinfo{person}{Fangai Liu}, {and} \bibinfo{person}{Weizhi Xu}.} \bibinfo{year}{2019}\natexlab{a}.
\newblock \showarticletitle{A survey on sequential recommendation}. In \bibinfo{booktitle}{\emph{2019 6th international conference on information science and control engineering (ICISCE)}}. IEEE, \bibinfo{pages}{106--111}.
\newblock


\bibitem[\protect\citeauthoryear{Yan, Cheng, Kang, Wan, and McAuley}{Yan et~al\mbox{.}}{2019}]%
        {yan2019cosrec}
\bibfield{author}{\bibinfo{person}{An Yan}, \bibinfo{person}{Shuo Cheng}, \bibinfo{person}{Wang-Cheng Kang}, \bibinfo{person}{Mengting Wan}, {and} \bibinfo{person}{Julian McAuley}.} \bibinfo{year}{2019}\natexlab{}.
\newblock \showarticletitle{CosRec: 2D convolutional neural networks for sequential recommendation}. In \bibinfo{booktitle}{\emph{Proceedings of the 28th ACM international conference on information and knowledge management}}. \bibinfo{pages}{2173--2176}.
\newblock


\bibitem[\protect\citeauthoryear{Yang}{Yang}{2019}]%
        {yang2019xlnet}
\bibfield{author}{\bibinfo{person}{Zhilin Yang}.} \bibinfo{year}{2019}\natexlab{}.
\newblock \showarticletitle{XLNet: Generalized Autoregressive Pretraining for Language Understanding}.
\newblock \bibinfo{journal}{\emph{arXiv preprint arXiv:1906.08237}} (\bibinfo{year}{2019}).
\newblock


\bibitem[\protect\citeauthoryear{Yao, Duan, Xu, Cai, Sun, and Zhang}{Yao et~al\mbox{.}}{2024}]%
        {yao2024survey}
\bibfield{author}{\bibinfo{person}{Yifan Yao}, \bibinfo{person}{Jinhao Duan}, \bibinfo{person}{Kaidi Xu}, \bibinfo{person}{Yuanfang Cai}, \bibinfo{person}{Zhibo Sun}, {and} \bibinfo{person}{Yue Zhang}.} \bibinfo{year}{2024}\natexlab{}.
\newblock \showarticletitle{A survey on large language model (llm) security and privacy: The good, the bad, and the ugly}.
\newblock \bibinfo{journal}{\emph{High-Confidence Computing}} (\bibinfo{year}{2024}), \bibinfo{pages}{100211}.
\newblock


\bibitem[\protect\citeauthoryear{Yuan, Yuan, Tan, Wang, Huang, and Huang}{Yuan et~al\mbox{.}}{2024}]%
        {yuan2024rrhf}
\bibfield{author}{\bibinfo{person}{Hongyi Yuan}, \bibinfo{person}{Zheng Yuan}, \bibinfo{person}{Chuanqi Tan}, \bibinfo{person}{Wei Wang}, \bibinfo{person}{Songfang Huang}, {and} \bibinfo{person}{Fei Huang}.} \bibinfo{year}{2024}\natexlab{}.
\newblock \showarticletitle{RRHF: Rank responses to align language models with human feedback}.
\newblock \bibinfo{journal}{\emph{Advances in Neural Information Processing Systems}}  \bibinfo{volume}{36} (\bibinfo{year}{2024}).
\newblock


\bibitem[\protect\citeauthoryear{Yuan, Duan, Tong, Shi, and Zhang}{Yuan et~al\mbox{.}}{2021}]%
        {yuan2021icai}
\bibfield{author}{\bibinfo{person}{Xu Yuan}, \bibinfo{person}{Dongsheng Duan}, \bibinfo{person}{Lingling Tong}, \bibinfo{person}{Lei Shi}, {and} \bibinfo{person}{Cheng Zhang}.} \bibinfo{year}{2021}\natexlab{}.
\newblock \showarticletitle{Icai-sr: Item categorical attribute integrated sequential recommendation}. In \bibinfo{booktitle}{\emph{Proceedings of the 44th international ACM SIGIR conference on research and development in information retrieval}}. \bibinfo{pages}{1687--1691}.
\newblock


\bibitem[\protect\citeauthoryear{Zhai}{Zhai}{2024}]%
        {zhai2024large}
\bibfield{author}{\bibinfo{person}{ChengXiang Zhai}.} \bibinfo{year}{2024}\natexlab{}.
\newblock \showarticletitle{Large language models and future of information retrieval: Opportunities and challenges}. In \bibinfo{booktitle}{\emph{Proceedings of the 47th International ACM SIGIR Conference on Research and Development in Information Retrieval}}.
\newblock


\bibitem[\protect\citeauthoryear{Zhang, Bao, Zhang, Wang, Feng, and He}{Zhang et~al\mbox{.}}{2023}]%
        {zhang2023chatgpt}
\bibfield{author}{\bibinfo{person}{Jizhi Zhang}, \bibinfo{person}{Keqin Bao}, \bibinfo{person}{Yang Zhang}, \bibinfo{person}{Wenjie Wang}, \bibinfo{person}{Fuli Feng}, {and} \bibinfo{person}{Xiangnan He}.} \bibinfo{year}{2023}\natexlab{}.
\newblock \showarticletitle{Is chatgpt fair for recommendation? evaluating fairness in large language model recommendation}. In \bibinfo{booktitle}{\emph{Proceedings of the 17th ACM Conference on Recommender Systems}}. \bibinfo{pages}{993--999}.
\newblock


\bibitem[\protect\citeauthoryear{Zhang, Gao, Zhang, Feng, Deng, and Hou}{Zhang et~al\mbox{.}}{2024a}]%
        {zhang2024ucl}
\bibfield{author}{\bibinfo{person}{Jing Zhang}, \bibinfo{person}{Hui Gao}, \bibinfo{person}{Peng Zhang}, \bibinfo{person}{Boda Feng}, \bibinfo{person}{Wenmin Deng}, {and} \bibinfo{person}{Yuexian Hou}.} \bibinfo{year}{2024}\natexlab{a}.
\newblock \showarticletitle{LA-UCL: LLM-augmented unsupervised contrastive learning framework for few-shot text classification}. In \bibinfo{booktitle}{\emph{Proceedings of the 2024 Joint International Conference on Computational Linguistics, Language Resources and Evaluation (LREC-COLING 2024)}}. \bibinfo{pages}{10198--10207}.
\newblock


\bibitem[\protect\citeauthoryear{Zhang, Yuan, Li, and Xu}{Zhang et~al\mbox{.}}{2024c}]%
        {zhang2024llm}
\bibfield{author}{\bibinfo{person}{Mingxuan Zhang}, \bibinfo{person}{Bo Yuan}, \bibinfo{person}{Hanzhe Li}, {and} \bibinfo{person}{Kangming Xu}.} \bibinfo{year}{2024}\natexlab{c}.
\newblock \showarticletitle{LLM-Cloud Complete: Leveraging cloud computing for efficient large language model-based code completion}.
\newblock \bibinfo{journal}{\emph{Journal of Artificial Intelligence General science (JAIGS) ISSN: 3006-4023}} \bibinfo{volume}{5}, \bibinfo{number}{1} (\bibinfo{year}{2024}), \bibinfo{pages}{295--326}.
\newblock


\bibitem[\protect\citeauthoryear{Zhang, Ladhak, Durmus, Liang, McKeown, and Hashimoto}{Zhang et~al\mbox{.}}{2024b}]%
        {zhang2024benchmarking}
\bibfield{author}{\bibinfo{person}{Tianyi Zhang}, \bibinfo{person}{Faisal Ladhak}, \bibinfo{person}{Esin Durmus}, \bibinfo{person}{Percy Liang}, \bibinfo{person}{Kathleen McKeown}, {and} \bibinfo{person}{Tatsunori~B Hashimoto}.} \bibinfo{year}{2024}\natexlab{b}.
\newblock \showarticletitle{Benchmarking large language models for news summarization}.
\newblock \bibinfo{journal}{\emph{Transactions of the Association for Computational Linguistics}}  \bibinfo{volume}{12} (\bibinfo{year}{2024}), \bibinfo{pages}{39--57}.
\newblock


\bibitem[\protect\citeauthoryear{Zheng, Hou, Lu, Chen, Zhao, Chen, and Wen}{Zheng et~al\mbox{.}}{2024}]%
        {zheng2024adapting}
\bibfield{author}{\bibinfo{person}{Bowen Zheng}, \bibinfo{person}{Yupeng Hou}, \bibinfo{person}{Hongyu Lu}, \bibinfo{person}{Yu Chen}, \bibinfo{person}{Wayne~Xin Zhao}, \bibinfo{person}{Ming Chen}, {and} \bibinfo{person}{Ji-Rong Wen}.} \bibinfo{year}{2024}\natexlab{}.
\newblock \showarticletitle{Adapting large language models by integrating collaborative semantics for recommendation}. In \bibinfo{booktitle}{\emph{2024 IEEE 40th International Conference on Data Engineering (ICDE)}}. IEEE, \bibinfo{pages}{1435--1448}.
\newblock


\bibitem[\protect\citeauthoryear{Zhou, Yu, Zhao, and Wen}{Zhou et~al\mbox{.}}{2022}]%
        {zhou2022filter}
\bibfield{author}{\bibinfo{person}{Kun Zhou}, \bibinfo{person}{Hui Yu}, \bibinfo{person}{Wayne~Xin Zhao}, {and} \bibinfo{person}{Ji-Rong Wen}.} \bibinfo{year}{2022}\natexlab{}.
\newblock \showarticletitle{Filter-enhanced MLP is all you need for sequential recommendation}. In \bibinfo{booktitle}{\emph{Proceedings of the ACM web conference 2022}}. \bibinfo{pages}{2388--2399}.
\newblock


\bibitem[\protect\citeauthoryear{Zhu, Yuan, Wang, Liu, Liu, Deng, Chen, Dou, and Wen}{Zhu et~al\mbox{.}}{2023}]%
        {zhu2023large}
\bibfield{author}{\bibinfo{person}{Yutao Zhu}, \bibinfo{person}{Huaying Yuan}, \bibinfo{person}{Shuting Wang}, \bibinfo{person}{Jiongnan Liu}, \bibinfo{person}{Wenhan Liu}, \bibinfo{person}{Chenlong Deng}, \bibinfo{person}{Haonan Chen}, \bibinfo{person}{Zhicheng Dou}, {and} \bibinfo{person}{Ji-Rong Wen}.} \bibinfo{year}{2023}\natexlab{}.
\newblock \showarticletitle{Large language models for information retrieval: A survey}.
\newblock \bibinfo{journal}{\emph{arXiv preprint arXiv:2308.07107}} (\bibinfo{year}{2023}).
\newblock


\end{thebibliography}

\balance

\end{document}